\documentclass{aastex631}

\shorttitle{\co\ and \cco\ in G148.04+0.11}
\shortauthors{Morgan et al.}

\begin{document}
\newcommand{\vdag}{(v)^\dagger}
\newcommand\aastex{AAS\TeX}
\newcommand\latex{La\TeX}
\newcommand{\tco}{$^{12}$CO}
\newcommand{\co}{$^{13}$CO}
\newcommand{\cco}{C$^{18}$O}
\newcommand{\degree}{$^{\circ}$}
\newcommand{\msol}{M$_{\odot}$}
\newcommand{\bigchi}{\raisebox{0.4ex}{\Large $\chi$}}

\title{A \co\ and \cco\ Study of the Planck Cold Clump G148.04+0.11 - probing the initial conditions of star cluster formation}

\correspondingauthor{Larry Morgan}
\email{lmorgan@nrao.edu}

\author[0000-0001-8708-5593]{Larry K. Morgan}
\affiliation{Green Bank Observatory \\
155 Observatory Rd
P.O. Box 2
Green Bank, WV 24944 (USA)}

\author[0000-0002-5881-3229]{David Eden}
\affiliation{Global Banking School, Universal Square, Devonshire Street North, Manchester, M12 6JH, UK}
\affiliation{Armagh Observatory and Planetarium, College Hill, Armagh, BT61 9DB, UK\\}

\author[0000-0003-1964-970X]{Chi-Yan Law}
\affiliation{INAF Osservatorio Astrofisico di Arcetri, Largo Enrico Fermi, 5, 50125 Firenze FI, Italy.}

\author[0000-0002-3389-9142]{Jonathan C. Tan}
\affiliation{Department of Physics \& Astronomy, Chalmers University of Technology, SE-412 96 Gothenburg, Sweden}
\affiliation{Department of Astronomy, University of Virginia, Charlottesville, VA 22904-4235, USA}

\begin{abstract}
We present new \co\ and \cco\ (J=1$\rightarrow$0) mapping observations of the Planck Galactic Cold Clump G148.04+0.11, obtained with the Argus array on the Green Bank Telescope. Using the $Astrodendro$ algorithm, we identify seven \co\ structures that separate into two kinematically distinct velocity groups, along with four denser \cco\ substructures associated with only three of the \co\ clumps. Adopting a distance of 1.327 ($\pm$ 0.042) kpc, we derive a total \co-traced mass of 651~\msol, with $Astrodendro$ leaf structures accounting for 280~\msol\ (43\%) of this total. A virial analysis of a compact and symmetric \cco\ core within one of these, labelled `b2 (core)', yields a virial ratio of 2.6 $\pm$ 0.7, consistent with this core being marginally gravitationally bound. Comparison with \tco\ data from the Exeter-FCRAO Galactic Plane Survey reveals a second, kinematically distinct velocity component along the line of sight, and we find no clear morphological correspondence between the \co/\cco\ structures and archival $Herschel$ dust continuum or embedded-source catalogues. Together, these results suggest that G148.04+0.11 is a cold, quiescent cloud without current star formation, in which the densest clumps are close to gravitational boundedness while the lower-density structures are unbound and/or pressure confined.

\end{abstract}


\section{Introduction}
\label{sec:intro}
The $Planck$ survey \citep{Planck2011} produced a catalogue of 13,188 $Planck$ Galactic Cold Clumps (PGCCs) \citep{PGCC2016}. These represent a unique population of cold, dense regions within the interstellar medium that are of particular interest to star formation studies. While PGCCs are home to conditions conducive to star formation (i.e. $T_d\sim$ 6 - 20 K), their YSO content is low, averaging half that of Infrared Dark Clouds (IRDCs) \citep{Eden2019}. PGCCs are not yet well-defined as a distinct class of objects and their relationship to other cold, dense structures in the Galaxy remains an areas of active research. They are thought to represent some of the earliest stages of star formation \citep{Wu2012,Liu2013,Yuebin2026}, potentially offering a probe of the initial conditions of protocluster clumps and pre-stellar cores.
This makes them valuable targets for studying the processes that initiate and regulate star formation on various scales. The overlap between PGCCs and IRDCs is significant, with many PGCCs also qualifying as IRDCs. However, PGCCs are generally colder and less dense than typical IRDCs, suggesting they may represent an earlier evolutionary stage. This continuum of properties from PGCCs to IRDCs to more active star-forming regions provides an opportunity to study the progression of star formation from its very beginnings.

PGCCs are found to be roughly on the threshold of virial stability, with 51\% of a sample of 96 cores found to be likely collapsing \citep{Zhang2016}. This marginal stability suggests that many PGCCs may be on the verge of initiating star formation, making them ideal laboratories for studying the triggers and early stages of this process. The filamentary nature of PGCCs \citep[e.g.,][]{Rivera-Ingraham2016,Kim2017} is consistent with current models of star formation, which emphasize the importance of filaments in the star formation process \citep{Andre2010,Andre2014}. PGCCs contain significant sub-structure, with 61\% of submm-detected sources having three or more compact sources \citep{Eden2019}. This hierarchical structure provides insights into how larger-scale cloud properties influence the formation of individual stars and clusters.

The object G148.04+0.11, first recorded in \citet{Planck2011} as a compact source, represents an excellent example of a PGCC that may be in the earliest stages of star formation. While not extensively studied, there have been previous observations made of G148.04+0.11 in $^{12}$CO, $^{13}$CO and C$^{18}$O \citep{Wu2012,Zhang2016} and HCO$^{+}$ and HCN \citep{Yuan2016}, although the properties of individual sources are not discussed. SCUBA-2 observations of G148.04+0.11 are also presented in \citet{Zhang2018} and in \citet{Eden2019} as part of the SCOPE project.

No previous study has resolved the structure of G148.04+0.11 in the molecular gas, or examined the properties of its individual substructures. In this paper we present new GBT/Argus \co\ and \cco\ (J=1$\rightarrow$0) maps of G148.04+0.11 at angular resolutions of 9\arcsec\ and 12\arcsec\ ($\sim$0.06 and 0.08 pc at our adopted distance of 1.327 kpc). Our goals are: (i) to identify and characterise the structure of the cloud, from the extended \co-traced gas to the denser condensations traced by \cco, (ii) to establish its kinematics, distance and relationship to other gas along the line of sight, (iii) to compare the molecular gas with dust continuum emission and embedded infrared sources and (iv) to derive the masses, abundances, turbulent motions and virial state of the cloud and its substructures in order to assess whether it shows evidence of gravitational collapse and star formation, or whether parts of it may instead be dispersing.

The paper is organised as follows. Section \ref{sec:obs} describes the observations and data reduction. In Section \ref{sec:data} we present the \co\ and \cco\ data, the $Astrodendro$ structure decomposition and the velocity structure of the cloud. Section \ref{sec:dist} discusses the distance to G148.04+0.11 and Section \ref{sec:FCRAO} uses data from the Exeter-FCRAO survey to place it in the context of other emission along the line of sight. Sections \ref{sec:Embedded} and \ref{sec:dust} compare the molecular gas with embedded sources and with dust continuum emission and Section \ref{sec:Environment} describes the wider environment of the cloud. In Section \ref{sec:Virial} we derive column densities, masses and abundances and analyse the virial state and turbulent motions of the gas. We discuss the results in Section \ref{sec:disc} and summarise our conclusions in Section \ref{sec:summary}.

\section{Observations and Data Reduction} \label{sec:obs}

Simultaneous on-the-fly mapping of the $^{13}$CO and C$^{18}$O emission (both in the J=1$\rightarrow$0 transition) was performed using the Argus instrument \citep{Sieth2014} on the Green Bank Telescope on the 13$^{th}$ of January 2020 under the project AGBT19B\_302. Argus is a 16-element heterodyne `radio camera' operating in the range of 85-116 GHz and is the product of a collaboration between Stanford U. (PI Sarah Church), Caltech, JPL, Univ. Maryland, Univ. Miami, and the Green Bank Observatory (GBO). The 16 heterodyne pixel receivers of Argus are mounted in a 4$\times$4 layout and have a beam separation of 30.4\arcsec\ in both the elevation and cross-elevation directions. 

Observations took place on the 13$^{\mathrm{th}}$ of January, 2020. The mapped region was set at a 12\arcmin\ square region, centred at $l$=148.051, $b$=+0.115. The Argus array size itself produces an undersampled margin to the final maps, resulting in a gridded area of $\sim$14\arcmin\ $\times$ 14\arcmin. A reference position of $l$=147.634, $b$=+0.597 was observed every 4$^{th}$ map row for the same duration as each scan row (35 s), after being checked for emission itself via pointed `Track' observations. The GBT maintains surface precision via `out-of-focus' holography observations, these were carried out towards 3C84 (NGC 1275) using the Ka-band receiver, resulting in an estimated surface accuracy of 229 microns. Pointing and focus observations were carried out at intervals of $\sim$50 min with measured pointing offsets of -2.1\arcsec\ -- +0.3\arcsec\ (mean = -0.4\arcsec) in azimuth and -0.8\arcsec\ -- +2.4\arcsec\ (mean = +1.0\arcsec) in elevation. The data were reduced using the GBTIDL package, with calibration achieved through observations of the Argus `Vane', essentially a warm load placed in the beam of the receiver. Atmospheric opacity estimates were taken from forecast values obtained from the CLEO\footnote{\url{https://www.gb.nrao.edu/~rmaddale/GBT/CLEOManual/index.html}} application, which are the average of the values from three surrounding local sites.

After baseline fitting and removal was performed on spectra from each ARGUS feed, these data were then gridded using the GBO tool `gbtgridder' with a Gaussian beam assumption and a pixel size of 2.75\arcsec. The resulting cubes were then subjected to further iterative baseline fitting and removal in order to equalize background noise levels. Under standard observing conditons, with a well-calibrated dynamic surface, the GBT would be expected to have a beamsize of $\sim$6.9\arcsec at these frequencies. However, a deliberate attempt was made to adopt the mapping/gridding technique described in, e.g., \citet{Holland1999} and (more fully) in \citet{Sawada2008}, for which the angle of scan direction to array axis is optimised to produce Nyquist sampling, despite less-than-Nyquist row-spacing. However, an error in the mapping parameter inputs used in this implementation resulted in the map being undersampled by a factor of $\sim$30\%. In order to compensate for the loss of spatial information, final FITS cubes were smoothed with a 2D Gaussian kernel such that the resulting angular resolution is 9\arcsec (FWHM).

The resulting data cubes have an rms noise of 0.2 K per pixel (on the T$_A^*$ scale) at a channel width of 0.09 km/s for both the \co\ and \cco\ (1–0) transitions. Peak \cco\ temperatures are low compared with this noise level. To increase the dynamic range in this transition, the \cco\ data were further spatially smoothed to a resolution of 12\arcsec, giving a mean rms noise of 0.14 K per pixel over the cube.

\section{Results and Analysis}
\begin{figure}
    \centering
    \includegraphics[width=\textwidth]{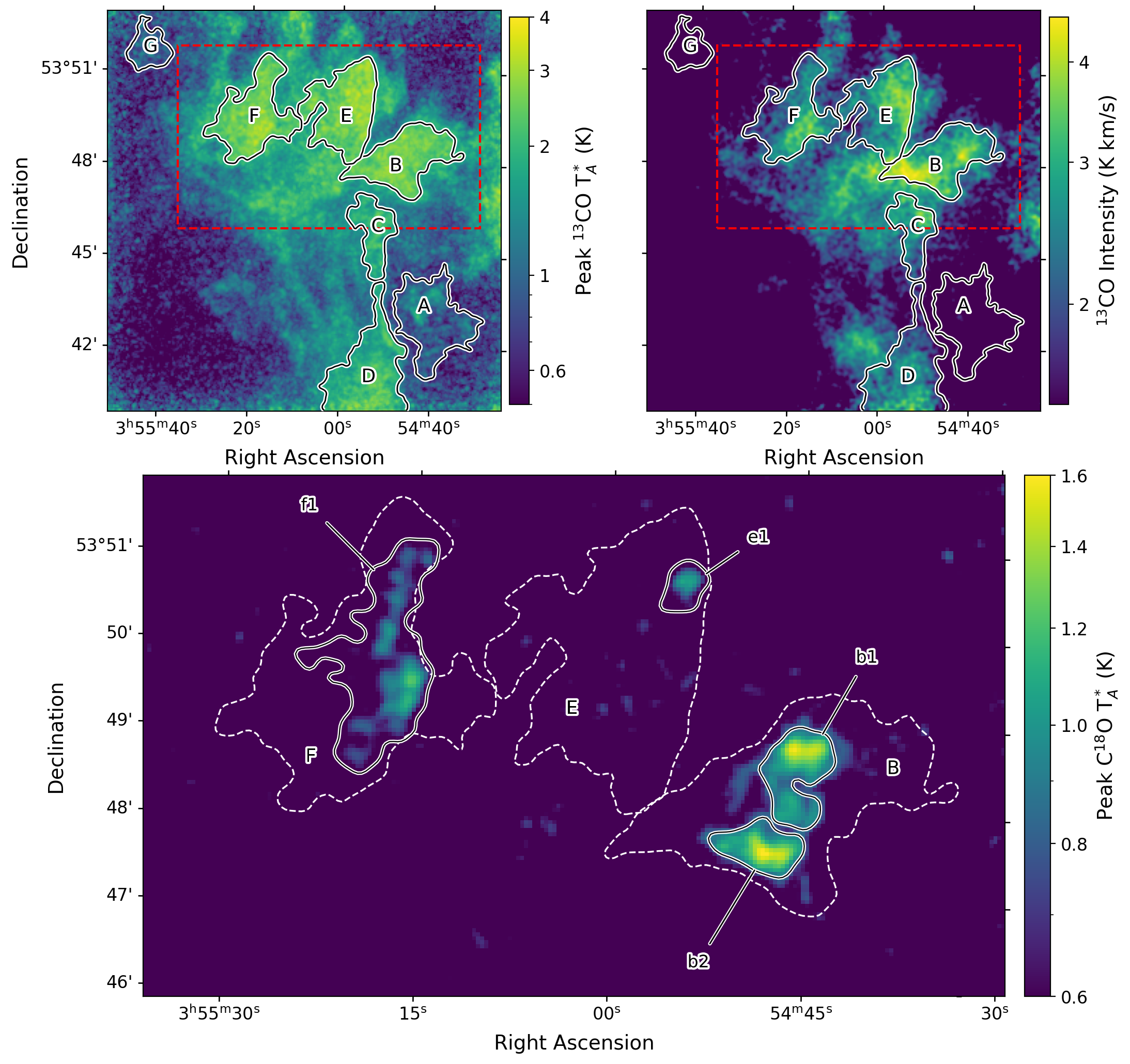}
        \caption{\textit{Top left:} \co\ peak corrected antenna temperature ($T_A^*$) map of G148.04+0.11, shown on a logarithmic colour scale. \textit{Top right:} \co\ integrated intensity (moment 0) map. Both top level images were produced over a velocity range of $-$8.31 to $-$3.73 km/s. Black-on-white contours outline the \co\ $Astrodendro$ leaf structures, labelled `A' to `G' in order of ascending R.A. The red dashed box marks the region shown in the lower panel. \textit{Lower panel:} \cco\ peak $T_A^*$ map of the boxed region, smoothed to a resolution of 12\arcsec\ and shown on a logarithmic colour scale. Black-on-white contours outline the \cco\ $Astrodendro$ leaf structures b1, b2, e1 and f1, as indicated. Dashed white contours show the parent \co\ structures B, E and F.}
    \label{fig:G148_COMaps}
\end{figure}

\subsection{\textsuperscript{13}CO and C\textsuperscript{18}O Data}
\label{sec:data}

Figure \ref{fig:G148_COMaps} (top left) shows the peak corrected antenna temperature ($T_A^*$) of the \co\ spectrum at each position across G148.04+0.11, with Fig. \ref{fig:G148_COMaps} (top right) showing the integrated intensity, or moment 0 map; both are derived from channels spanning $-$8.31 to $-$3.73 km/s, chosen to encompass the full extent of \co\ emission associated with G148.04+0.11 (see Section \ref{sec:FCRAO}) while excluding the unrelated, diffuse component near $-$1.4 km/s. These images suggest stronger `clumpy' structure towards the upper half, while weaker emission towards the lower middle has a filamentary appearance. In order to robustly define these structures, the $Astrodendro$ package \citep{Robitaille2019} was used. This package implements the algorithm of \citet{Rosolowsky2008}, establishing a hierarchical tree-like sequence of structures, leafs, branches and trunks\footnote{note that only the `leaf' and `branch' nomenclature was identified by \citet{Rosolowsky2008}, here we use `trunks' to identify the most basic structure found in our $Astrodendro$ analysis}.
This is achieved via the selection of three main parameters, $min\_value$, the minimum voxel value required for inclusion in a structure, $min\_delta$, the minimum voxel value difference required to separate distinct structures and $min\_npix$, the minimum number of voxels required to define an independent structure. Here we adopt values of $min\_value$ = 3$\sigma$, where $\sigma$ is the mean standard deviation in our spectra (0.2 K), $min\_delta$ = 5$\sigma$ and $min\_npix$ = 1050.  These parameters reflect the observational sensitivity and spatial scale of the region, and therefore do not have obvious `default' values. Our choices aim to capture the broader emission features while minimizing spurious detections and avoiding an excessive number of independent structures, particularly given the relatively small field of view. We assume a minimal linewidth of 0.2 km/s, representative of starless cores \citep{Tafalla2004} and corresponding to the non-thermal linewidth expected for typical coherent structures \citep{Goodman1998}. This is also approximately the sound speed defining thermally critical filaments \citep{Arzoumanian2013}. This thus corresponds to a FWHM of 0.47 km/s, or five channels at our velocity resolution of $\sim0.09$ km/s (Section \ref{sec:obs}).

In order to determine the impact of a spatial coverage threshold, we performed a series of $Astrodendro$ analyses. This began with a minimum spatial coverage of a single beam area, equivalent to $\sim$8.4 pixels (i.e. $\pi(\theta_{\rm FWHM}/2)^2$ for $\theta_{\rm FWHM}$ = 9\arcsec\ and 2.75\arcsec\ pixels), across the minimum of five velocity channels described above. Since $min\_npix$ is a count of voxels, this corresponds to $min\_npix$ = 8.4 $\times$ 5 $\approx$ 42. We then increased this threshold iteratively in integer multiples of the beam area. As might be expected, increasing $min\_npix$ was found to progressively remove the smallest, least coherent structures. From examination, a threshold of approximately 25 beamsizes is found to be appropriate for our dataset, excluding only structures found on the edge of our input image, in comparison to the single beamsize results. Edge pixels are subject to lower sensitivity, due to the multi-feed focal plane array nature of our observations. Subsequently, while `real' emission is present on the edges of our image, here we exclude small, marginal structures at the edges, while retaining larger, coherent regions which have bulk emission well within the mapped area.

\begin{figure}
    \centering
    \includegraphics[width=0.7\textwidth]{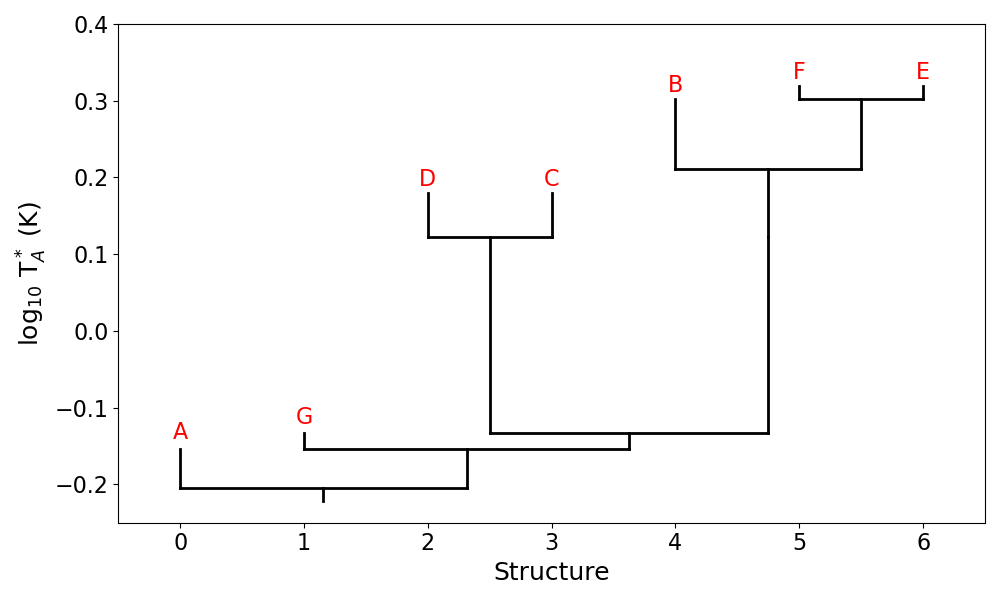}
    \caption{$Astrodendro$ structure diagram. The `leaf' structures outlined in Fig. \ref{fig:G148_COMaps} are labelled.}
    \label{fig:G148_13CO_ADStructure}
\end{figure}

The contours overlaid on the panels of Fig. \ref{fig:G148_COMaps} show the results of this analysis, with the peak \co\ $T_A^*$ map of Fig. \ref{fig:G148_COMaps} (top left) set on a logarithmic colour scale. Black-on-white contours outline the `leaf' structures found by the $Astrodendro$ analysis. A total of seven structures are found, all of which are superficially flat and broad in nature. These have been labelled `A' through `G' in order of ascending R.A.. Figure \ref{fig:G148_13CO_ADStructure} shows the overall structure of G148.04+0.11, as determined by $Astrodendro$. It can be seen that all of the structures labelled `A' through `G' in Fig. \ref{fig:G148_COMaps} are emergent from the overall low-level broad emission of the cloud. Further, we see that the `A' and `G' structures are relatively weak in emission and separate from the other structures at the lowest levels. Structures `B' through `F' are relatively stronger in emission and form two separate branches of the dendrogram. `C' and `D' are grouped together on one branch, while `B', `E' and `F' form another, with `E' and `F' grouped together, separate from `B'. 

The \cco\ emission detected in our observations of this region is not widespread at levels above the noise threshold. As described in Section \ref{sec:obs}, smoothing to a final image resolution of 12\arcsec\ has been performed in order to enhance weaker emission where possible, while maintaining a balance with the high angular resolution of the original data. Figure \ref{fig:G148_COMaps} (lower) shows the peak corrected antenna temperature from these smoothed \cco\ data, for the region marked by the red dashed box in Fig. \ref{fig:G148_COMaps} (upper panels). We have deliberately restricted this panel to exclude large areas in which no emission was detected. The relevant mask contours from the \co\ $Astrodendro$ analysis are overlaid as dashed white contours and labelled, showing that \cco\ emission is associated only with the `B', `E' and `F' clumps. It can be seen that there is widespread \cco\ emission over the lowest longitude/R.A. region of these, identified here as clump B, while clumps E and F contain significantly lower obvious \cco\ emission. Clump E exhibits only a small condensation to the north-west of the clump and clump F showing emission along the western edge of the `F' clump contour.

The \cco\ data was analysed using $Astrodendro$, with input parameters similar but notably different to our \co\ analysis. For the \cco\ cube we again use a $min\_value$ of 3$\sigma$ but use a $min\_delta$ value of 3$\sigma$, compared to 5$\sigma$ for the \co\ analysis. This was determined due to the contrasting goals of the respective analyses. In the case of our \co\ data, our main goal is to identify broad-scale emission and trace material throughout the full extent of our map. In the case of the \cco\ data, we do not see large-scale emission and further wish to identify smaller condensations in any case. \cco\ is a tracer of dense gas and thus highlights potential dense cores which may be associated with active collapse and/or star formation. It should be noted that the $\sigma$ value used in our \cco\ analyses (0.15 K) is slightly lower than for our \co\ data due to the additional smoothing of the \cco\ cube (see Section \ref{sec:obs}). A further difference between our \co\ and \cco\ analyses is that the $min\_pix$ value used for the \cco\ analysis corresponds to the size of a single beam. This is due to the insensitivity of the $Astrodendro$ results to this parameter in the case of the \cco\ data. A total of five structures were found by $Astrodendro$, where four of these are `leafs', with two of those emerging from a single branch. These leaf structures are shown in Fig. \ref{fig:G148_COMaps} (lower) and are largely as described above - clump `F' (as identified from the \co\ analysis) shows \cco\ emission along its western edge and clump `E' exhibits a small condensation to the north-west. Clump `B' contains two distinct leaf structures. Visual inspection of Fig. \ref{fig:G148_COMaps} (lower) suggests that the northernmost leaf could plausibly be further divided into two separate components. We explored this possibility by varying the input parameters of $Astrodendro$. Although such a subdivision could be produced, the two resulting components remained connected at a lower hierarchical level, emerging from a common branch that was effectively equivalent to the structure shown in Fig. \ref{fig:G148_COMaps} (lower). Moreover, adopting these parameter settings led to the identification of several additional structures that were consistent with noise rather than physically meaningful emission. We therefore retain the original decomposition and label the \cco\ clumps in relation to the sources found from the \co\ $Astrodendro$ analysis as `b1','b2' `e1' and `f1'.

A comparison of the three panels of Fig. \ref{fig:G148_COMaps} highlights the clumpy structure present in the upper half of the observed region and the filamentary appearance of the material seen in \co\ (but not \cco) within the lower half. These features are typical of early-stage star-forming regions: Filaments have recently been recognised as widespread features in molecular clouds, playing a crucial role in the star formation process \citep{Andre2010, Molinari2010, Andre2014}, while the loosely described `clumpy' structure often represents the initial condensations that may eventually form stars \citep{Bergin2007, Johnstone2006, Motte1998}. 

\begin{figure}
    \centering
    \includegraphics[width=0.8\textwidth]{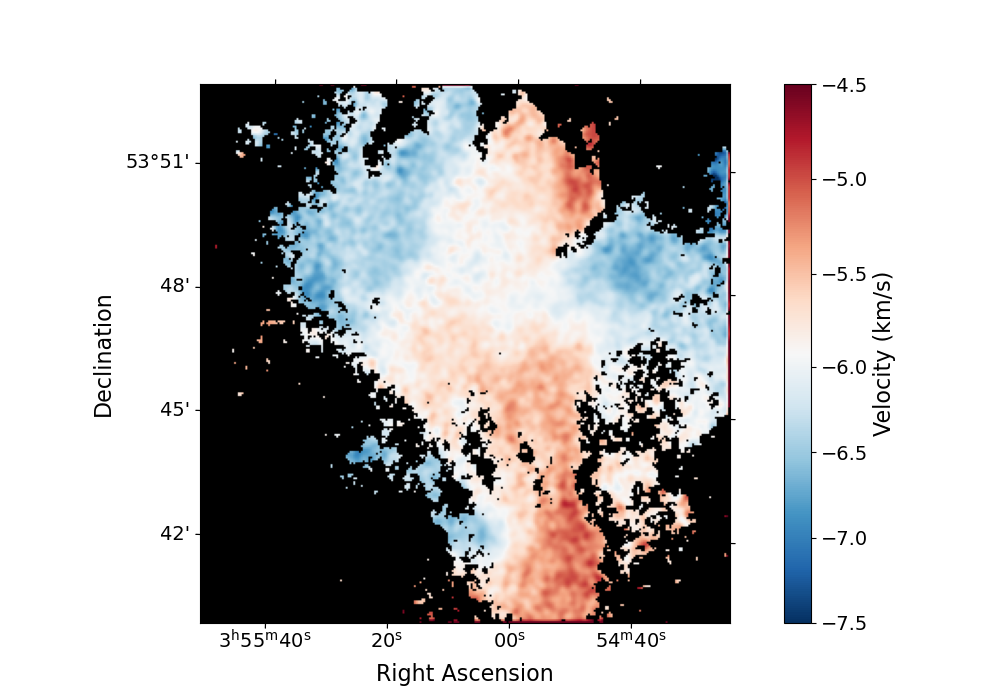}
    \caption{$^{13}$CO moment 1, velocity map of G148.04+0.11. This image has been smoothed with a Gaussian kernel of width ($\sigma$) 1 pixel and masked to pixels for which the peak \co\ line temperature exceeds a value of five times the spectral RMS noise for that pixel. The velocity field is displayed using a diverging colormap symmetrically normalised about the median velocity value of -5.9 km/s to emphasize deviations from this median velocity. }
    \label{fig:G148_13CO_M1}
\end{figure}

\begin{figure}
    \centering
    \includegraphics[width=\textwidth]{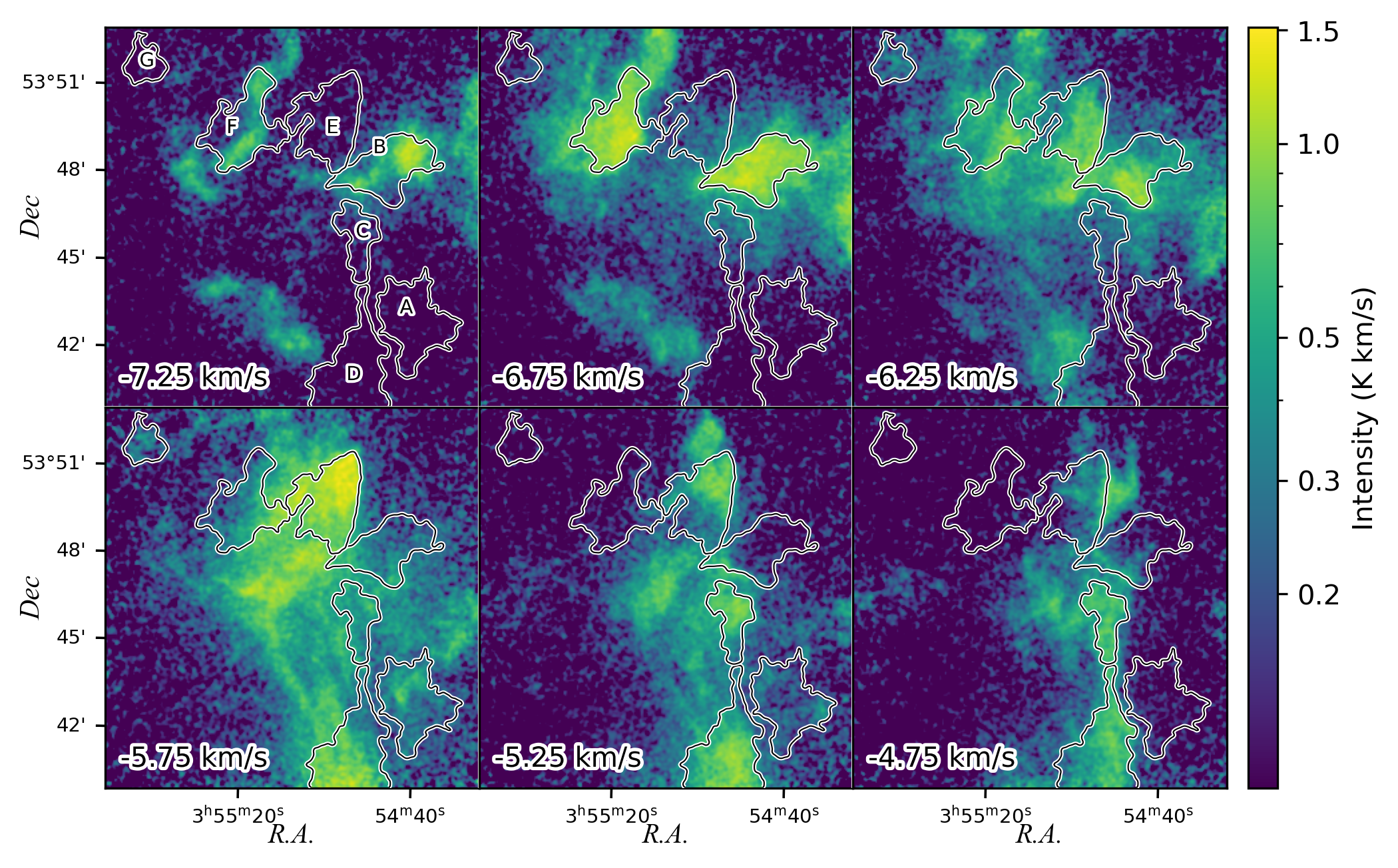}
    \caption{\co\ integrated channel maps of G148.04+0.11 over velocity intervals of 0.5 km/s, with the central velocity of each interval given in the lower left corner of each panel. The $Astrodendro$ leaf structures `A' to `G' are outlined in each panel and labelled in the first.}
    \label{fig:G148_13CO_ChanMap}
\end{figure}

Investigating these indications of structure further, Fig.\ref{fig:G148_13CO_M1} shows the moment 1, velocity map of the region. This image has been masked to pixels for which the peak \co\ line temperature exceeds a value of five times the spectral RMS noise for that pixel. It has also been smoothed with a Gaussian kernel of 1 pixel width (resulting in an overall spatial angular resolution of 11\arcsec). This was done in order to highlight and enhance any broader scale attributes across the region. The lower bound for the size of star-forming clumps is assumed to be $\sim$0.1 pc (see e.g. \citet{Myers1983}, \citep{Seale2012} and references within). The spatial scale of a single 2.75\arcsec\ pixel in Fig.\ref{fig:G148_13CO_M1} is equivalent to $\sim$0.01 - 0.03 pc (depending on the estimated distance of the cloud, see Section \ref{sec:dist}), with a resolution limit spatial equivalent of 0.04 - 0.12 pc. Therefore, we do not expect to have obscured any potentially visible coherent velocity features via this process. It can be seen that the velocity of material over this region ranges from $\sim$\ -7.0 to -4.5 km/s. An initial, simplistic interpretation of this image is that there are two regions, one towards the north-west of the image and one towards the north-east, that exhibit velocities lower than $\sim$\ -5.5 km/s. Equally, a broad feature running north to south across the region exhibits velocities greater than this mid-range value. In order to investigate this apparent contrast, we present channelised segments of our \co\ emission cube with contours of the $Astrodendro$ structures overlaid in Fig. \ref{fig:G148_13CO_ChanMap}. Each image panel shows \co\ emission integrated over a velocity width of 0.5 km/s, with the central velocity of each interval labelled in the lower left corner of each panel. The top left image represents a `near' velocity of -7.25 km/s, velocity increasing by 0.5 km/s per frame from left to right across two rows to the `far' image at -4.75 km/s.

Figure \ref{fig:G148_13CO_ChanMap} shows that there is a coherent velocity structure encompassing all of the structures identified by the $Astrodendro$ process, although this appears to be separated into two kinematically distinct regions within the cloud. Overall, the \co\ emission spans a range of approximately -7.25 to -4.75 km/s. At velocities around -6.75 km/s our structures `B' and `F' dominate, while `E' is conspicuously devoid of emission. This suggests a distinct kinematical offset of structure `E' to `B' and `F', despite their spatial proximity. Around -6.25 km/s, these structures all show emission, although it is diffuse by comparison to lower velocities. Some emission is seen related to the structures `A' and `G' between velocities of -5.25 and -6.25 km/s (with `A' being correlated with slightly redder emission and `G' bluer). This emission is notably diffuse, consistent with the description made earlier and suggesting lower column density material related to these structures.

At -5.75 km/s, structure `E' dominates the emission and is clearly related to a larger filamentary structure with a north/south orientation that was not recovered as an independent structure. Or, rather, the large-scale structure forms a `branch' in our $Astrodendro$ analysis which encompasses the broader region underlying all of the `leafs' except for `A' and `G'. Structures `C' and `D' are the dominant `leaf' structures at the reddest cloud velocities, -5.25 to --4.75 km/s.

Overall, the observed velocities suggest a broad division of G148.04+0.11 into two groups. At the blue end of the range the clumps `B' and `F'  peak near -6.75 km/s, while `E' peaks near -5.75 km/s. In the red end, the more filamentary structures `C' and `D' peak near -5.0 km/s. Structures `A' and `G' represent more diffuse, intermediate-velocity material and are considered separate to these groupings. The velocity separation of approximately 1.5–2.0 km/s between the groups may indicate distinct sub-cloud components or reflect large-scale velocity gradients within the molecular cloud.

\subsection{Cloud Distance}
\label{sec:dist}
The distance to the molecular gas under consideration is not readily apparent. The peak gas velocity is fairly narrowly constrained to a range of -7.0 to -4.5 km/s. However, this region of the Galaxy is not necessarily well understood in terms of kinematic models. \citet{Reid2009, Reid2014,Reid2019} measured trigonometric parallaxes and proper motions for masers associated with high-mass star-forming regions throughout the Milky Way. The results of those studies indicate strong support for a four-arm spiral structure to the Milky Way and, using the derived model, a kinematic distance estimation can be obtained\footnote{\url{http://bessel.vlbi-astrometry.org/revised_kd_2014}}. Adopting a systemic velocity of $-$6.0 km/s for G148.04+0.11, corresponding to the peak of the associated EXFC \co\ emission (Section \ref{sec:FCRAO}) and consistent with the median velocity of $-$5.9 km/s in our Argus \co\ moment 1 map, and a velocity uncertainty of 5 km/s \citep{Reid2019}, this calculator provides a distance estimate of 0.23$^{+0.41}_{-0.23}$ kpc. The original catalogue identification of this cloud (PGCC G148.03+0.10) \citep{PGCC2016} assigned a distance of 0.77 kpc, based on models of near-infrared extinction. \citet{Zhang2018} used the Bayesian distance calculator of \citet{Reid2016} to estimate a distance of 2.15 kpc. However, this estimate was based on a velocity range of -32.49 to -34.31 km/s, corresponding to Perseus Arm emission along this line of sight (Section \ref{sec:FCRAO}) rather than the $-$6.0 km/s gas of G148.04+0.11. \citet{Guo2021}, using a method independent of gas kinematics derived a distance of 1.327 ($\pm$ 0.042) kpc by applying a simple reddening model to using the stellar catalogues of \citet{Chen2019} and \citet{Green2019}. We used the $dustmaps$ Python package \citep{Green2018} in conjunction with the 3D dustmaps of \citet{Edenhofer2024} and \citet{Green2019} to evaluate extinction values along the line of sight to G148.04+0.11 and found results consistent with those of \citet{Guo2021}, we therefore continue with an assumption of 1.327 kpc.

\begin{figure}
    \centering
    \includegraphics[width=1.0\textwidth]{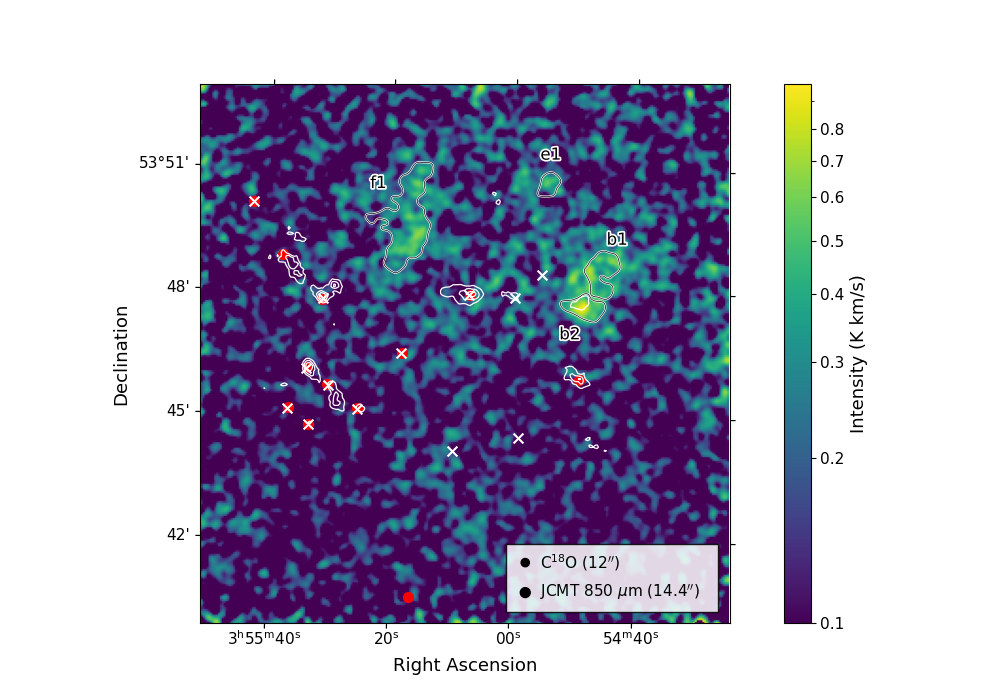}
    \caption{\cco\ integrated emission map of G148.04+0.11, with $Herschel$ HPPSC sources at wavelengths of 70 and 160 $\mu$m represented by red circles and white x symbols, respectively. White contours show emission at 850 \micron\ with levels of 5, 8, 11 and 14 times the image rms, observed as part of the SCOPE project \citep{Eden2019}. The beam sizes of the \cco\ and 850 \micron\ observations are depicted in the lower right corner.}
    \label{fig:C18O_Embedded}
\end{figure}

\subsection{Exeter-FCRAO CO Galactic Plane Survey Data}
\label{sec:FCRAO}
\begin{figure}
    \centering
    \includegraphics[width=1.0\textwidth]{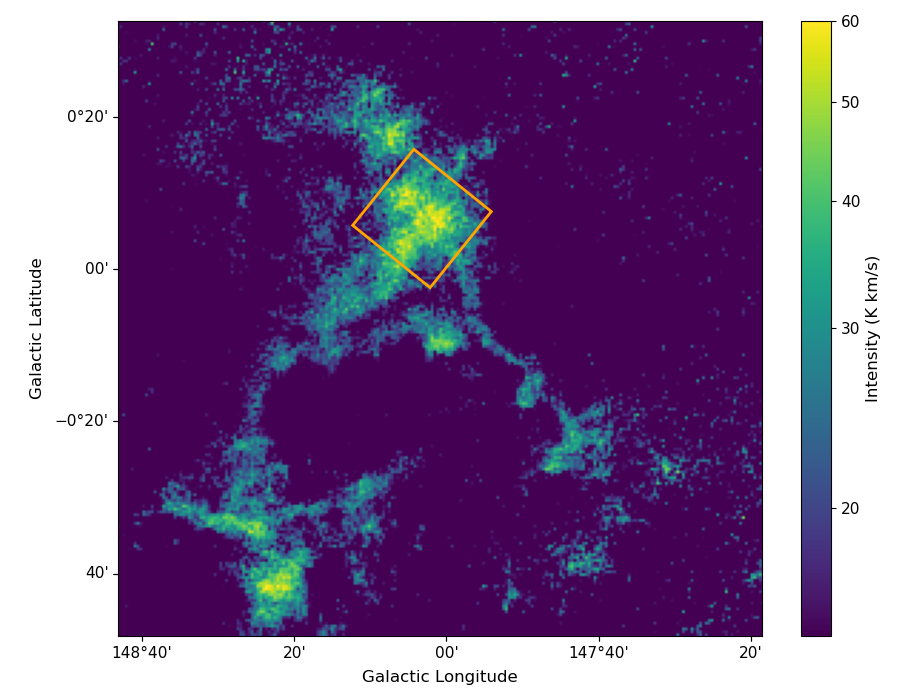}
    \caption{\tco\ emission, as measured by the EXFC  \citep{Mottram2010}, integrated over a velocity range of -42.5 km/s $<$ V$_{LSR}$ $<$ 8.0 km/s. The overlaid orange square shows the region of the observations presented here.}
    \label{fig:GRS_12CO}
\end{figure}

\begin{figure}
    \centering
    \includegraphics[width=1.0\textwidth]{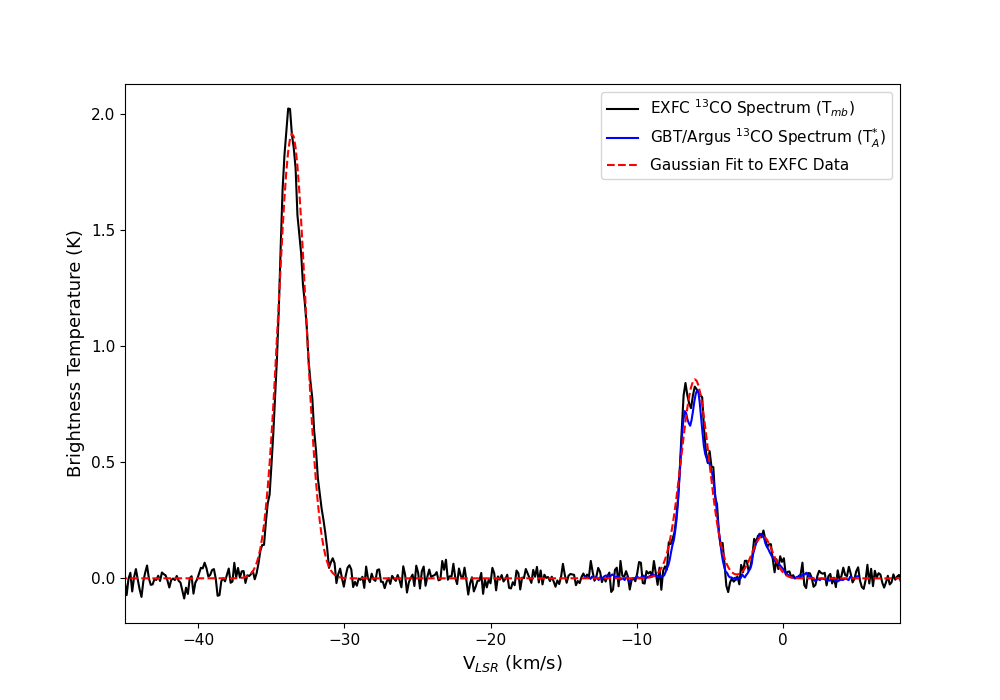}
    \caption{\co\ spectral emission, as measured by the EXFC  \citep{Mottram2010}, averaged over the region of the observations presented here. Overlaid in blue is the average spectrum taken from the Argus map observed in this work. The red dashed line shows lines of best-fit Gaussian functions, resulting in V$_{LSR}$ estimates of  -33.6, -6.0 and -1.4 km/s.}
    \label{fig:GRS_13COSpec}
\end{figure}

The direction towards which we have performed our observations passes through the Local Arm as well as the Perseus Arm \citep{Reid2019}.
In order to more robustly investigate the velocity associations of the gas detected in our presented observations, we have considered the \tco\ and \co\ line emission from the Exeter-FCRAO CO Galactic Plane Survey (EXFC: \citealt{Mottram2010}; C. M. Brunt et al., in preparation).  While the EXFC data comparatively have a considerably poorer angular resolution of 50\arcsec\ in beamwidth, with a single-pixel RMS sensitivity of 0.7 K, as a survey dataset there is naturally a far greater coverage of Galactic longitude and latitude available. Further, while the data from our own observations encompass a velocity range of approximately -13.5 to 5 km/s, the EXFC data at this Galactic longitude cover a far wider range of approximately -105 to 22 km/s. Figure \ref{fig:GRS_12CO} shows integrated \tco\ emission of the EXFC covering the broader region surrounding G148.04+0.11. This emission has been integrated over the velocity range -42.5 km/s $<$ V$_{LSR}$ $<$ 8.0 km/s, the extent of any discernible emission at this location. Averaged over the region observed in this study, the EXFC \tco\ and \co\ spectra both show three strong line components. Figure \ref{fig:GRS_13COSpec} shows the \co\ spectral emission as extracted from the EXFC with the spectral emission from our presented observations overlaid. We have further fit Gaussian components to each observed peak from which we derive V$_{LSR}$ values of  -33.6, -6.0 and -1.4 km/s.

\begin{figure}
    \centering
    \includegraphics[width=0.49\textwidth]{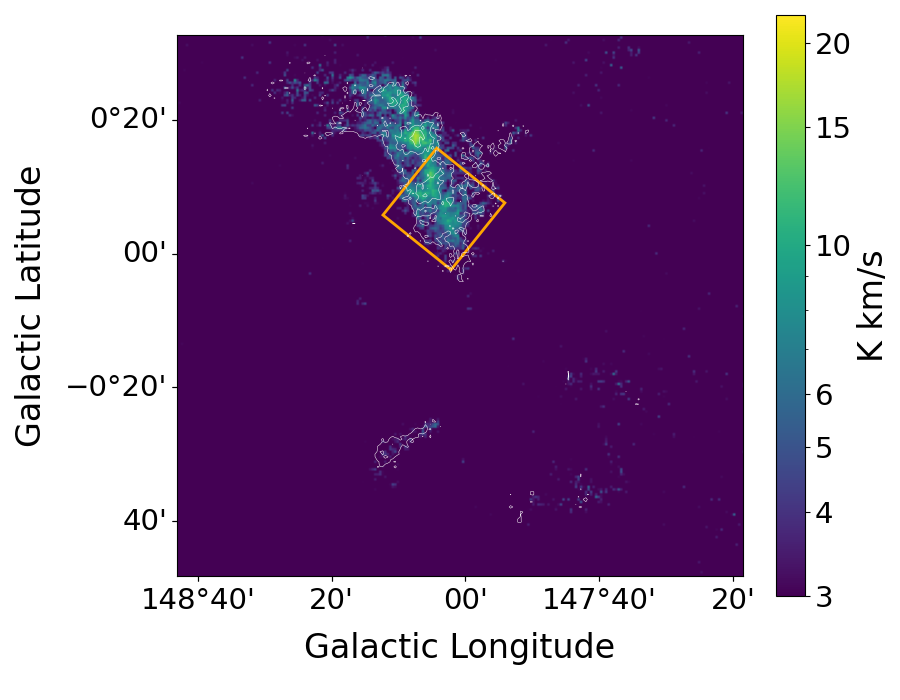}
    \includegraphics[width=0.49\textwidth]{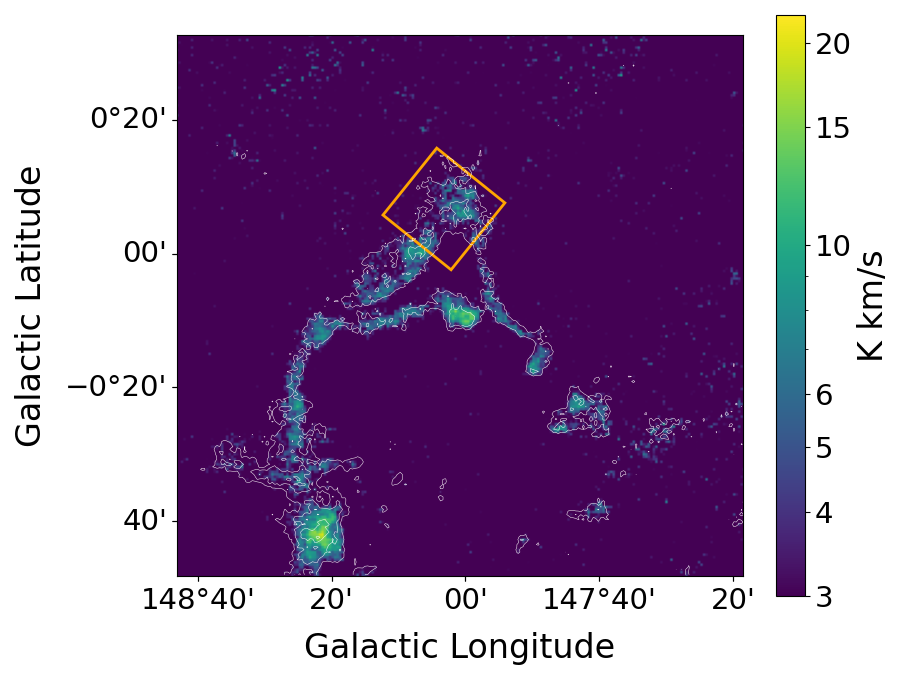}
    \caption{\co\ emission, as measured by the EXFC  \citep{Mottram2010}, integrated over a velocity range of -40 km/s $<$ V$_{LSR}$ $<$ -28.0 km/s (left, Perseus Arm component) and -12 km/s $<$ V$_{LSR}$ $<$ 5.0 km/s (right, low-velocity component, containing G148.04+0.11). White contours show \tco\ emission integrated over the same velocity ranges at levels of 30, 50 and 75\% of the relevant peak value. The overlaid orange square shows the region of the observations presented here.}
    \label{fig:GRS_13CO}
\end{figure}

We present images of the \co\ emission as measured by the EXFC, integrated over velocity ranges of -12 - 5 km/s and -40 - -28 km/s, in Fig. \ref{fig:GRS_13CO}. Applying the Bayesian distance calculator of \citet{Reid2019} to the $-$33.6 km/s peak seen in Fig. \ref{fig:GRS_13COSpec} places this gas in the Perseus Arm at a distance of 2.05 $\pm$ 0.22 kpc, consistent with the 2.15 kpc estimate of \citet{Zhang2018}, and $\sim$0.7 kpc beyond G148.04+0.11 at our adopted distance of 1.327 kpc.
Based on these results, we hereafter refer to these two velocity ranges as a low-velocity component (containing G148.04+0.11 at $-$6.0 km/s as well as the weaker $-$1.4 km/s feature) and the Perseus Arm component, respectively.
It is evident that our mapped region incorporates gas observed in both of the velocity ranges we have identified. However, there is a clear distinction between the two velocity ranges in terms of the broader visible structure. In the Perseus Arm gas (-40 km/s $<$ V$_{LSR}$ $<$ -28.0 km/s) image there is an obvious extension of the \co\ emission towards greater Galactic latitudes but very little towards lower latitudes.  Conversely, the low-velocity (-12 km/s $<$ V$_{LSR}$ $<$ 5.0 km/s) image has very limited emission towards higher Galactic latitudes but shows distinct structure at lower latitudes. The relevance of these structures to our own observations is limited, given the limited velocity range and spatial extent of our own observations. However, the presence of Perseus Arm gas in our surveyed region should be noted when examining dust continuum emission.

\subsection{Embedded Sources}
\label{sec:Embedded}
In order to investigate the star-forming potential of the observed cloud, a search was performed for embedded far-infrared/submillimeter sources. The $Herschel$ Hi-GAL compact source (band-merged) catalogue \citep{Elia2021} does not contain any sources in the observed region. However, a search of the ESA Herschel Science Archive\footnote{\url{http://archives.esac.esa.int/hsa/whsa/}} returned 18 sources from the $Herschel$/PACS Point Source Catalogue (HPPSC; \citealt{Marton2017}), extracted with SUSSEXtractor and aperture photometry at a threshold of 3$\sigma$. These sources are listed in Table \ref{tab:Herschel}. Fifteen of these sources have counterparts in the Hi-GAL single-band catalogues, extracted with CuTEx \citep{Molinari2011} following the procedure of \citet{Molinari2016} and used to construct the 360$^{\circ}$ catalogue of \citet{Elia2021}. The ten sources detected at both 70 and 160 \micron\ have colour temperatures of 20–28 K, typical of Galactic dust heated by embedded young stellar objects. Contamination from extragalactic sources is not expected for any but the three faintest sources found, based on the galaxy number counts of \citet{Bethermin2010} (integrated over our 0.047 deg$^2$ field). Figure \ref{fig:C18O_Embedded} shows integrated emission of \cco\ from our observations with overlaid contours taken from the JCMT/SCUBA-2 850 \micron\ continuum observations of \citet{Eden2019}. Red circles and white `x' symbols show the positions of $Herschel$ HPPSC sources at wavelengths of 70 and 160 $\mu$m, respectively. It can be seen that, while the 850 \micron\ emission broadly matches the positions of the HPPSC sources, these sources do not correlate with regions of strong \cco\ emission, nor even the brighter regions of \co\ integrated intensity (cf. Fig. \ref{fig:G148_COMaps}, top right and lower). One small region of 850 \micron\ emission does correspond to the strongest peak of \cco\ integrated intensity (`b2') at R.A. = 03:54:48.8, Dec = 53:47:38.

\begin{table*}
\centering
\footnotesize
\begin{tabular}{clcccccll}
\hline\hline
ID & HPPSC name & R.A. & Dec. & $S_{70}$ (S/N) & $S_{160}$ (S/N) & $T_{\rm col}$ & \multicolumn{2}{c}{Hi-GAL counterpart} \\
 & & (J2000) & (J2000) & (Jy) & (Jy) & (K) & 70\,$\mu$m & 160\,$\mu$m \\
\hline
1 & J035449.0+534555 & 03:54:49.0 & +53:45:55 & 0.22 (5.8) & \ldots & \ldots & 148.0251+0.0767 & 148.0247+0.0766$^a$ \\
2 & J035455.4+534828 & 03:54:55.4 & +53:48:28 & \ldots & 0.24 (3.3) & \ldots & \ldots & \ldots \\
3 & J035458.9+534430 & 03:54:58.9 & +53:44:30 & \ldots & 0.85 (8.4) & \ldots & \ldots & 148.0586+0.0739 \\
4 & J035459.8+534753 & 03:54:59.8 & +53:47:53 & \ldots & 0.52 (5.9) & \ldots & \ldots & 148.0243+0.1188 \\
5 & J035507.1+534757 & 03:55:07.2 & +53:47:57 & 0.40 (11.0) & 1.44 (11.3) & 21--23 & 148.0372+0.1312 & 148.0374+0.1315 \\
6 & J035509.6+534409 & 03:55:09.7 & +53:44:09 & \ldots & 0.16 (3.8) & \ldots & \ldots & \ldots \\
7 & J035516.5+534036 & 03:55:16.5 & +53:40:36 & 0.07 (3.6) & \ldots & \ldots & \ldots & \ldots \\
8 & J035518.1+534631 & 03:55:18.2 & +53:46:31 & 0.20 (6.7) & 1.02 (12.4) & 20--22 & 148.0732+0.1304 & 148.0740+0.1302 \\
9 & J035525.4+534508 & 03:55:25.4 & +53:45:08 & 0.57 (16.0) & 2.02 (10.1) & 21--23 & 148.1017+0.1240 & 148.1019+0.1237 \\
10 & J035530.3+534543 & 03:55:30.3 & +53:45:43 & 1.60 (25.7) & 2.52 (6.4) & 25--28 & 148.1048+0.1388 & 148.1049+0.1387 \\
11 & J035531.4+534747 & 03:55:31.5 & +53:47:47 & 1.81 (40.4) & 5.27 (19.1) & 22--24 & 148.0849+0.1676 & 148.0848+0.1675 \\
12 & J035533.3+534445 & 03:55:33.3 & +53:44:45 & 0.34 (9.5) & 0.72 (5.5) & 24--26 & 148.1207+0.1313 & 148.1211+0.1316 \\
13 & J035533.7+534607 & 03:55:33.8 & +53:46:07 & 1.43 (17.3) & 4.43 (9.0) & 22--24 & 148.1073+0.1497 & 148.1074+0.1496 \\
14 & J035536.6+534508 & 03:55:36.7 & +53:45:08 & 0.23 (5.7) & 0.99 (8.7) & 21--22 & 148.1232+0.1414 & 148.1236+0.1417 \\
15 & J035537.9+534851 & 03:55:38.0 & +53:48:51 & 2.12 (41.7) & \ldots & \ldots & 148.0859+0.1914 & 148.0868+0.1900$^a$ \\
16 & J035542.9+535008 & 03:55:43.0 & +53:50:08 & 2.30 (53.5) & 3.63 (11.4) & 25--28 & 148.0815+0.2156 & 148.0819+0.2157 \\
17 & J035545.5+534841$^b$ & 03:55:45.5 & +53:48:41 & 0.27 (4.7) & \ldots & \ldots & 148.1019+0.2009 & 148.1028+0.2010$^a$ \\
18 & J035545.5+535106$^b$ & 03:55:45.5 & +53:51:06 & 0.28 (6.3) & 1.17 (8.2) & 21--23 & 148.0761+0.2323 & 148.0761+0.2315 \\
\hline
\end{tabular}
\begin{minipage}{\textwidth}
\vspace{4pt}
\footnotesize
\caption{$Herschel$/PACS Point Source Catalogue sources within the observed region. Positions are from the 70\,$\mu$m entry where available. $T_{\rm col}$ is the 70/160\,$\mu$m colour temperature for a modified blackbody with $\beta$ = 2.0--1.5. Hi-GAL counterparts (designations G$\ell$$\pm$$b$) are from the Hi-GAL single-band compact source catalogues, extracted with CuTEx \citep{Molinari2011} following \citet{Molinari2016} and used for the 360\degree\ catalogue of \citet{Elia2021}; matched within 2\arcsec\ unless noted.\\
$^a$\,Not detected at 160\,$\mu$m in the HPPSC; the listed Hi-GAL 160\,$\mu$m counterpart lies 0.2\arcsec\ (ID 1), 5.4\arcsec\ (ID 15) or 3.2\arcsec\ (ID 17) from the 70\,$\mu$m position.\\
$^b$\,Within $\sim$20 pixels of the eastern map edge, where sensitivity is reduced.}
\label{tab:Herschel}
\end{minipage}
\end{table*}
\subsection{Correlation with Continuum/Dust Emission}
\label{sec:dust}
\begin{figure}
    \centering
    \includegraphics[width=1.0\textwidth]{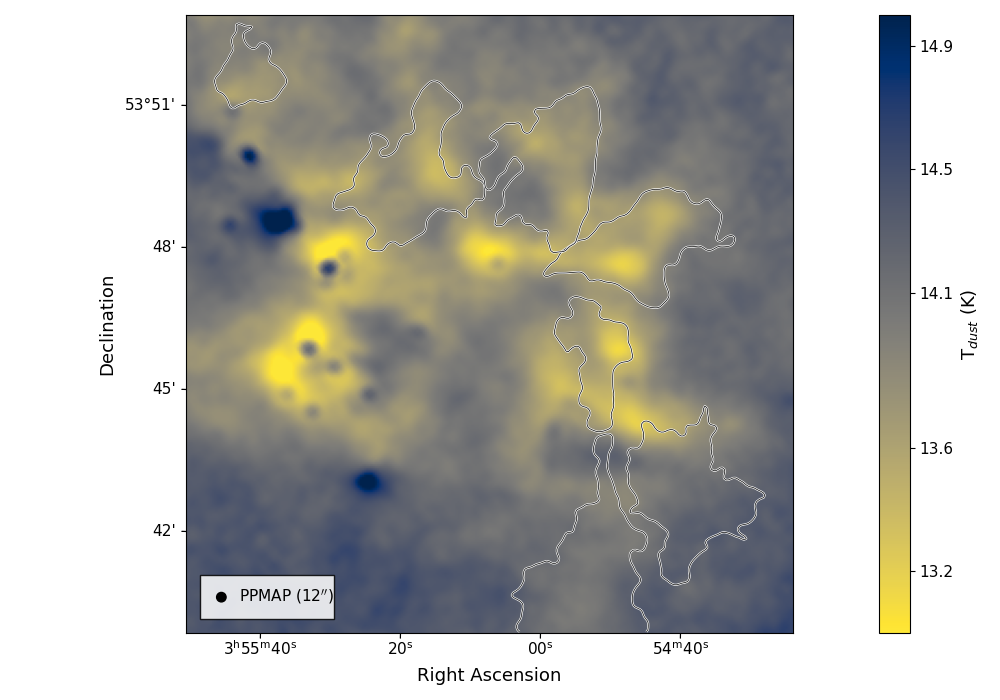}
    \caption{$Herschel$ Hi-GAL PPMAP dust temperature image, contours show the location of the \co\ structures identified using $Astrodendro$. The filled circle (lower left) shows the 12\arcsec\ effective angular resolution of the PPMAP maps \citep{Marsh2017}.}
    \label{fig:PPMAPTd}
\end{figure}

\begin{figure}
    \centering
    \includegraphics[width=1.0\textwidth]{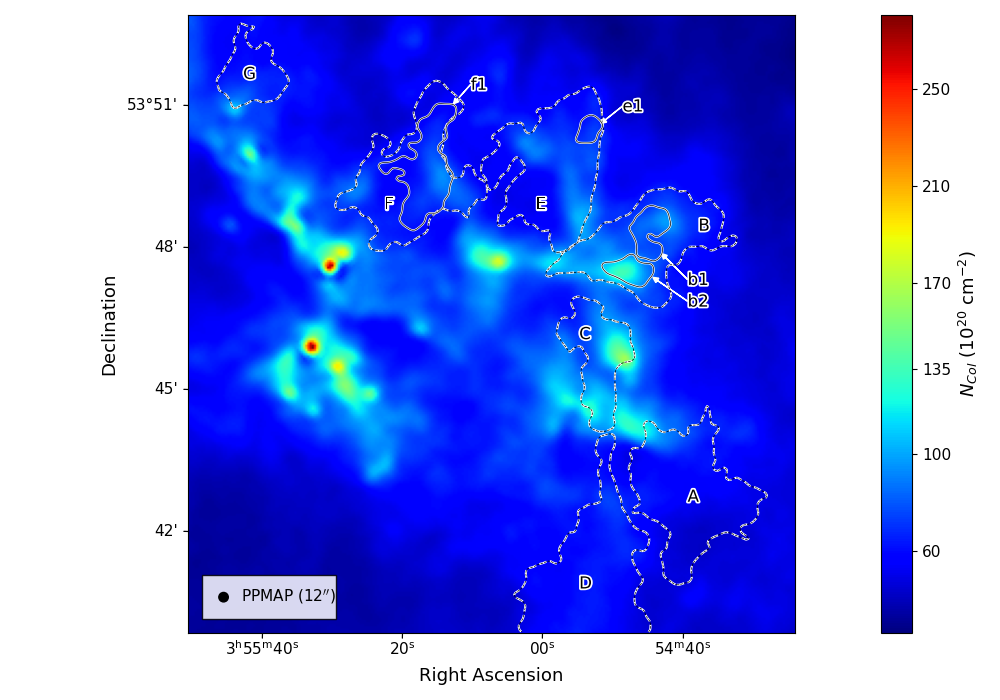}
    \caption{$Herschel$ Hi-GAL PPMAP column density image, dashed contours show the location of the \co\ structures identified using $Astrodendro$, while solid contours show the \cco\ structures. The filled circle (lower left) shows the 12\arcsec\ effective angular resolution of the PPMAP maps \citep{Marsh2017}}
    \label{fig:PPMAPNcol}
\end{figure}

\citet{Marsh2017} presented column density maps of the Galactic plane, derived using the PPMAP procedure\footnote{available at \url{http://www.astro.cardiff.ac.uk/research/ViaLactea/}}. These maps are based upon $Herschel$ Hi-GAL \citep{Molinari2010} continuum data over the range 70 - 500 \micron\ and allow for determinations of dust-derived column density, with an angular resolution of 12\arcsec. Figures \ref{fig:PPMAPTd} and \ref{fig:PPMAPNcol} show the column density and temperature determinations for the regions mapped in this work. The column density maps are evaluated at dust temperatures ranging from 8.0, to 50 K, with uniform spacing in log T$_{D}$. The final column density map are summations of the individual evaluations at each dust temperature, with averaged dust temperature maps the result of weighting obtained from the fractional column densities. 

Inspection of the individual dust temperature maps reveals that the \co\ structures identified using $Astrodendro$ (outlined in Figs  \ref{fig:PPMAPTd} and \ref{fig:PPMAPNcol}) are strongly associated with cold dust. Specifically, none of the structures `A' - `G' have a column density contribution of greater than 1\% of their total integrated column density arising from the PPMAP slices above 18.4 K. We therefore have eliminated the contributions above this threshold from Fig. \ref{fig:PPMAPNcol} in order to maintain clarity - nearby sources described in Section \ref{sec:Embedded} dominate the column density scaling in the image at higher temperatures.

Figure \ref{fig:PPMAPNcol} shows that there is little morphological correlation between our structures and the observed dust emission as traced by the column density map. Possible exceptions are a compact emission feature coincident with the `C' structure, a column density enhancement coinciding with \cco\ emission towards the west of the `f1' structure and emission which appears likely related to `b1' and `b2' (particularly `b2'). There is emission within the contours of the `E' and `G' structures, although the emission associated with `E' is spatially offset from the `e1' \cco\ contour, and the emission within the `G' contour appears more likely to be associated with an elongated structure running northeast-southwest in the image.

The compact feature located within the `C' structure is clearly associated with the HiGAL 70 \micron\ source noted in Section \ref{sec:Embedded}, which also exhibits 850 \micron\ emission. Notably no \cco\ emission is detected at this location and, while there is a modest \co\ peak near this position (at an angular separation of 0.61\arcmin - 0.24 pc at the assumed distance of 1.327 kpc), it does not exhibit either the intensity nor morphology which might be expected in the presence of an embedded 70 \micron\ source. The feature is centred at a position of 03:54:49.4, +53:45:52.9 (J2000), with EXFC \co\ emission present at -5.1 km/s, although there is notably stronger emission seen at -34.5 km/s. The angular resolution of the EXFC is insufficient to distinguish any morphological similarities between the emission at these velocities and the dust emission. The \co\ peak temperature at this local peak is 2.57 K, compared to 1.7 K at the position of the column density peak. Estimating the \co/\cco\ abundance ratio directly at this location is difficult given the absence of significant \cco\ emission; however, ratios derived from the `b1', `b2', `e1' and `f1' regions range from 6.0 to 16.0 (see Section \ref{sec:Virial}). Applying these ratios to the observed \co\ temperatures yields predicted \cco\ peak temperatures of 0.16 - 0.43 K (at the \co\ emission peak) and 0.11 - 0.28 K (at the local column density peak). Given that the spectral RMS of our \cco\ data is $\sim$0.15 K, these predicted values are at or below our detection threshold. While we tentatively argue that the presence of 850 \micron\ emission, along with an embedded 70 \micron\ source should provide an easily detectable level of \cco\ emission and that, therefore, this source is likely located at a significantly different position along the line-of-sight, we are unable to state this conclusively.

The spatial offset between the `e1' \cco\ structure and the nearest column density enhancement, taken to be centred at 03:55:02.4, +53:50:15.6 (J2000), is 1.27\arcmin. At the assumed distance of 1.327 kpc, this corresponds to a projected physical separation of 0.49 pc. Inspection of EXFC \co\ data at the position of the dust enhancement reveals emission peaks at both -5.7 km/s and -34.5 km/s, consistent with contributions from both the G148.04+0.11 cloud and the more distant Perseus Arm component. The same dual-velocity signature is present at the position of `e1' itself. Our \cco\ data do not cover the velocity range necessary to investigate any potential emission at -34.5 km/s.

The column density enhancement seen at the edge of the `G' structure is centred at approximately 03:55:45.3, +53:51:03.3 (J2000). EXFC \co\ data at this position show emission only at -33.7 km/s, with no counterpart at the systemic velocity of the G148.04+0.11 cloud. This strongly suggests that the dust emission in this region is associated with the Perseus Arm component and is unrelated to the G148.04+0.11 cloud.

The brightest peak of \cco\ seen in our data - within the `b2' $Astrodendro$ mask - has a visible counterpart in the HiGAL column density image, as does the swath of emission associated with `f1'. However, other regions, which are bright in dust emission and/or are coincident with the locations of 70/160 $\mu$m sources, do not show obvious correlation with strong emission in either \co\ or \cco\ from our observations. Given that our observations cover a Galactic latitude range of $\sim$-0.038\degree\ to 0.27\degree, it is likely that line-of-sight confusion exists towards this region and that the observed dust emission, along with the noted 70 and 160 \micron\ sources, is tracing a region at a significant distance from the gas covered by our observations. As we found with the EXFC data (Section \ref{sec:FCRAO}), we are evidently observing two regions separated by $\sim$30 km/s in velocity and $\sim$0.7 kpc along the line of sight: G148.04+0.11 at 1.33 kpc and the Perseus Arm at $\sim$2.05 kpc.
 
\subsection{Broader Environment}
\label{sec:Environment}
The location of G148.04+0.11 coincides with multiple previously identified features which have potential relevance to the formation and development of the material observed there. Towards lower latitudes there is a well-established large H$_{II}$ region, LBN 148.11-0.45 \citep{Lynds1965}, originally included by \citet{Sharpless1959} as part of Sh2-205. \citet{Dobashi2005} used Digitized Sky Survey I\footnote{\url{https://www.ipac.caltech.edu/doi/irsa/10.26131/IRSA441}} data to identify dark clouds based on $A_v$ measurements based on star counts. Cloud 942 of that work is extensive, covering Galactic longitude/latitude ranges of 8.5\degree and 4.9\degree, respectively. Within this cloud, 20 local peaks were identified as `clumps' and, of these, the clump P7 is located at $l$=148.07,$b$=+00.10. This clump has an extent of 0.17\degree x 0.43\degree ~and is centred close to the central point of the \co ~map presented here. A later study \citep{Dobashi2011}  performed a similar search at higher resolution, using the 2 Micron All Sky Survey Point Source Catalog \citep{Cutri2003}. While \citet{Dobashi2005} aimed to trace the global extent of dark clouds and diffuse regions, \citet{Dobashi2011} had the goal of identifying dense cloud cores leading to star formation. From this study, we see two of the \citet{Dobashi2011} dark clouds (4040 and 4041) within our mapped region, although neither of these are correlated with peaks of \co\ or \cco\ emission, nor with IR/submillimetre emission at 70, 160 or 850 \micron\ (see Fig.\ref{fig:C18O_Embedded}).

\subsection{Virial Analysis}
\label{sec:Virial}
In order to determine the current relevance of the observed region in the context of potential star formation, as well as establishing an informed projection of what the fate of this region might be, we wish to establish the opacity and column density of the gas. From this, we may establish an estimate of the mass and use this to identify any potential star formation. 

We determine the column density of \cco\ via
\begin{equation}
N_{col} = \frac{8\pi\nu^{3}}{c^{3}A_{ul}}\,\frac{Q_{\rm rot}}{g_{u}}\,
\frac{e^{E_{u}/k_{B}T_{ex}}}{e^{h\nu/k_{B}T_{ex}}-1}\,
\frac{\int T_{b}\,dv}{J(T_{ex})-J(T_{bg})},
\qquad
Q_{\rm rot} \simeq \frac{k_{B}T_{ex}}{hB}\,e^{hB/3k_{B}T_{ex}}
\label{eg:Ncol}
\end{equation}
\citep[e.g.][]{Mangum2015}, which assumes that the \cco\ emission is optically thin and that the level populations are in local thermodynamic equilibrium (LTE) at a single excitation temperature, $T_{ex}$. Here the Einstein coefficient, $A_{ul}$ = 6.30 $\times$ 10$^{-8}$ s$^{-1}$ and $g_u$ = 3 and $E_u/k_B$ = 5.27 K, the degeneracy and energy of the upper ($J_u$ = 1) level of the \cco\ $J$ = 1$\rightarrow$0 transition, respectively. The rest frequency of this transition is $\nu$ = 109.7821734 GHz. $A_{ul}$ is calculated from the permanent dipole moment, $\mu$ = 0.11079 Debye. The rotational partition function, $Q_{\rm rot}$, uses the approximation of \citet{McDowell1988}, where $B$ is the rotational constant (54891.420 MHz for \cco). $J(T)$ is equal to $\frac{h \nu}{k_B} \left[e^\frac{h \nu}{k_B T}-1 \right]^{-1}$ and $T_b$ is the brightness temperature, determined here assuming a main beam efficiency for the GBT at 110 GHz of 0.51. This is derived via peak observations of the source 3C345, using a flux value of 2.77 Jy taken from the ALMA calibrator page\footnote{\url{https://almascience.eso.org/sc/}}.

Eq. \ref{eg:Ncol} requires the excitation temperature of \cco, which cannot be determined from the (largely) optically thin \cco\ line alone. We therefore assume LTE, in which all CO isotopologues share a common $T_{ex}$. Deriving $T_{ex}$ from a line's own peak brightness temperature, by inverting $T_b = J(T_{ex}) - J(T_{bg})$, requires the emission to be optically thick ($\tau \gg 1$) and to fill the beam. Neither condition is reliably met for \co\ or \cco, which both have $\tau \lesssim 1$ (see below). Their clumpy structure also means that their beam-filling factors may be less than unity. Applying this inversion to \co\ and \cco\ within the \cco\ structures gives median values of only 8.1 and 4.2 K, respectively, which are therefore lower limits. We instead derive $T_{ex}$ on a pixel-by-pixel basis from the peak brightness temperature of the optically thick \tco\ line in the EXFC data (Section \ref{sec:FCRAO}), using the same relation. \tco\ traces more extended, lower-density gas than its isotopologues and we therefore assume a beam-filling factor of unity. However, as the EXFC data have an angular resolution of 50\arcsec, emission structured on smaller scales will be diluted within the beam, so that these values of $T_{ex}$ may be underestimated. Within the \cco\ structures, the resulting $T_{ex}$ has a median of 14.0 K, in agreement with the dust temperatures of 13--15 K derived from PPMAP (Section \ref{sec:dust}). An independent estimate can be made from the \co/\cco\ line ratio, but only where \cco\ is detected at high signal-to-noise, which is limited to the \cco\ structures identified in Section \ref{sec:data}. There, assuming an abundance ratio of 7.55 \citep{Wilson1994}, the line ratio gives a median $T_{ex}$ of 9.1 K. This lower value should be treated with caution as it depends on the assumed abundance ratio and is biased low due to the selection effects related to restricting the analysis to pixels with high \cco\ signal-to-noise. Because $N_{col}$ depends weakly on $T_{ex}$ in this regime, adopting the lower value would reduce the \cco\ column densities by $\sim$13\%. Adopting the \tco-derived $T_{ex}$, the \cco\ optical depth (Eq. \ref{eq:tau}) within the \cco\ structures has a median of 0.16, confirming that \cco\ is largely optically thin, while the \co\ optical depth within the \co\ structures has a median of 0.53. We have chosen not to apply the optical depth correction of \citet{Goldsmith1999} (Eq. \ref{eq:Goldsmith})to \cco\ here, though note that doing so would increase the \cco\ column densities by $\sim$8\% in the median ($\lesssim$15\% in 95\% of pixels) and lower the \co/\cco\ abundance ratios derived below by a similar fraction. As a result these ratios should be regarded as upper limits. The \cco-based masses are unaffected, as they are scaled to the \co\ column density via the pixel-by-pixel abundance ratio.

The column density of \co\ is found by applying a correction to Eq. \ref{eg:Ncol} for emission that is not optically thin \citep{Goldsmith1999}. This correction is valid for any optical depth and tends to unity in the optically thin limit:
\begin{equation}
N^{thick}_{col} = N^{thin}_{col} \frac{\tau}{1-e^{-\tau}}
\label{eq:Goldsmith}
\end{equation}

Here, naturally, it is necessary to find the optical depth, $\tau$ of the \co\ transition. This is determined via 
\begin{equation}
\tau=-\ln\left(1-\frac{T_{b}}{J(T_{ex})-J(T_{bg})}\right)
\label{eq:tau}
\end{equation}

Figure \ref{fig:Ncol} shows the log of \co\ column density, overlaid with contours of \cco\ (log) column density. It can be seen that the \co\ and \cco\ column densities generally trace the structure seen in the integrated intensity map of \co\ (Fig. \ref{fig:G148_COMaps}, top right) and the \cco\ peak temperature map (Fig. \ref{fig:G148_COMaps}, lower), as might be expected.

We convert column density to mass via
\begin{equation}
M = \frac{\mu_{H_2}\, m_H\, N_{col}\, \theta}{\mbox{$\bigchi$}}
\end{equation}
where $\mu_{H_2}$ = 2.8 is the mean molecular weight per hydrogen molecule, assuming 1 He for every 5 $\rm H_2$ molecules, $m_H$ is the mass of a hydrogen atom (1.673 x 10$^{-27}$ kg), $\bigchi$ is the abundance ratio of the relevant molecule and $\theta$ is the area of the region being evaluated. We find values of $\bigchi$ for \co\ via EXFC \tco\ moment 0 emission, in conjunction with the formulation of \citet{Bolatto2013},  $N(\mathrm{H_2}) = W(\mathrm{^{12}CO})$ $\times$~2 $\times$~10$^{20}$ cm$^{-2}$ with $\pm$30\% uncertainty for $N(\mathrm{H_2})/W(\mathrm{^{12}CO})$.  Integrating emission over the low-velocity range of -12 - 5 km/s (Section \ref{sec:FCRAO}) and taking the ratio of the derived $N(\mathrm{H_2}$) value with our \co\ column density leads to a range of \co\ abundance, (0.82 - 4.67) $\times$~10$^{-6}$, with a mean of 1.68 $\times$~10$^{-6}$. These are consistent with values found in the fully-shielded regime across Orion A and Orion B, (1.4 - 3.4) $\times$~10$^{-6}$ \citep{Ripple2013}, as well as independent values of 2.5 $\pm$1.5  $\times$~10$^{-6}$ found in Orion B by \citet{Roueff2021}. All of these values fall well below those found by \citet{Pineda2008} for regions in Perseus, where they report 2.8 $\times$~10$^{-5}$ $<$ \co/H$_2$ $<$ 4.9 $\times$~10$^{-5}$.

Abundance values of \cco\ were determined via ratios of column density maps, smoothed to a 12\arcsec\ resolution. Due to the lack of widespread detectable \cco\ emission, this was evaluated only over the regions identified as `b1', `b2', `e1' and `f1' (see Fig. \ref{fig:G148_COMaps}, lower). A further threshold mask was applied in which only regions above 3$\sigma$ in the M0 map were considered. Evaluating these, we find the \co/\cco\ ratio in these regions ranges from 6.0 - 16.0, with a mean of 9.4; as noted above, these are upper limits, since no optical depth correction has been applied to the \cco\ column densities. This is broadly consistent with the value of 7.55 predicted by \citet{Wilson1994} via their formulation of $\raisebox{0.4ex}{{$\bigchi$}$^{13}_{18}$}$  = $\frac{58.8 R_{GC} +37.1}{7.5 R_{GC} +7.6}$, where $R_{GC}$ is the galactocentric distance of our source, 9.33 kpc.

The mass of the entire observed region determined from our \co\ observations, above a three sigma moment 0 threshold, is calculated at 651~\msol. The individual regions identified by $Astrodendro$ have masses as shown in Table \ref{tbl:Masses}. The uncertainties associated with the presented masses combine, in quadrature, the statistical noise in the integrated \co\ intensity (accounting for correlated noise within the beam) and the 27\% pixel-to-pixel scatter in the derived \co\ abundance. These uncertainties are conservative, as the abundance scatter itself includes some observational noise. However, we further note that an additional systematic error of 31\% applies to all mass estimates, due to uncertainties in the CO-to-H$_2$ conversion factor ($\pm$30\%; \citealt{Bolatto2013}) and distance ($\pm$6\%).

\begin{table}[h]
\centering
\begin{tabular}{ |c|c|c|c|c|c| }
\hline
Structure & Mass & $R_{\rm eff}$ & $\sigma_{obs}$ & $M_v$ & $\alpha_{vir}$ \\
 & (\msol) & (pc) & (km/s) & (\msol) & \\
 \hline
\multicolumn{6}{|c|}{\co} \\
 \hline
A & 17.1 $\pm$ 4.9 & 0.38 & 0.46 & 112 & 6.5 $\pm$ 1.9 \\
B & 81.5 $\pm$ 22.1 & 0.43 & 0.49 & 142 & 1.7 $\pm$ 0.5 \\
C & 14.0 $\pm$ 3.9 & 0.21 & 0.51 & 73 & 5.2 $\pm$ 1.5 \\
D & 39.2 $\pm$ 10.7 & 0.36 & 0.59 & 162 & 4.1 $\pm$ 1.1 \\
E & 64.7 $\pm$ 17.6 & 0.40 & 0.49 & 131 & 2.0 $\pm$ 0.6 \\
F & 60.4 $\pm$ 16.4 & 0.43 & 0.42 & 113 & 1.9 $\pm$ 0.5 \\
G & 3.1 $\pm$ 1.1 & 0.19 & 0.39 & 41 & 13 $\pm$ 5 \\
\hline
\multicolumn{6}{|c|}{\cco} \\
\hline
b1 & 9.5 $\pm$ 2.7 & -- & -- & -- & -- \\
b2 & 9.6 $\pm$ 2.7 & -- & -- & -- & -- \\
b2 (core) & 4.0 $\pm$ 1.1 & 0.09 & 0.23 & 10.3 & 2.6 $\pm$ 0.7 \\
e1 & 3.3 $\pm$ 1.0 & -- & -- & -- & -- \\
f1 & 27.9 $\pm$ 7.7 & -- & -- & -- & -- \\
\hline
\end{tabular}
\caption{Masses associated with each of the \co\ and \cco\ $Astrodendro$ structure masks. For the \co\ and `b2 (core)' structures, the effective radius, observed velocity dispersion, virial mass and virial ratio are also given (see text). The `b2' row gives the full $Astrodendro$ mask; the `b2 (core)' row gives the compact core within it analysed in Section \ref{sec:Virial}.}
\label{tbl:Masses}
\end{table}

As referenced in Section \ref{sec:Embedded} and seen in Figs \ref{fig:C18O_Embedded} and \ref{fig:Ncol}, there is a distinct condensed region of emission visible in \cco\ towards the lower end of the `B' structure. This source is contained within the `b2' structure identified by our \cco\ dendrogram analysis, though we note that the full `b2' $Astrodendro$ mask extends beyond the region of compact, symmetric emission and submillimetre association. We distinguish between these two structures here, and in Table \ref{tbl:Masses}, as `b2' and `b2 (core)'. The apparent circular symmetry of `b2 (core)' and its association with (weak) submillimetre emission, may indicate the potential for collapse and star-forming activity. A 2D Gaussian fit to the \cco\ integrated intensity over this region yields beam-deconvolved standard deviations of 0.097 and 0.084 pc for $\sigma_{maj}$ and $\sigma_{min}$, respectively. This is a beam-deconvolved aspect ratio of 1.16, consistent with approximate circular symmetry. Adopting this radius as appropriate for a virial analysis of this source, we adopt a threshold of 56\% of the peak integrated intensity, which encloses radius of 0.091~pc. The intensity-weighted velocity dispersion of \cco\ over this aperture, computed as the quadrature sum of the mean dispersion from Gaussian fits to the individual spectra (see below) and the pixel-to-pixel variation of the fitted centroid velocities, is 0.23~km/s. The total one-dimensional velocity dispersion of the gas, $\sigma_{tot}$, is obtained by removing the thermal contribution of the \cco\ molecule and adding that of the mean gas particle, $\sigma_{tot}^2 = \sigma_{obs}^2 - k_BT_{ex}/m_{C^{18}O} + k_BT_{ex}/(\mu m_H)$. Using the mean \tco-derived excitation temperature over the core, $T_{ex}$ = 14.3 K, the thermal dispersions are 0.06 and 0.22~km/s, respectively, giving $\sigma_{tot}$ = 0.31~km/s. The virial mass of a uniform-density sphere is then $M_v = 5R\sigma_{tot}^{2}/G$ \citep{Bertoldi1992}, equivalent to the homogeneous-sphere case of \citet{MacLaren1988}, giving $M_v$ = 10.3~\msol\ for $R$ = 0.091 pc.

In order to assess the dynamical state of the gas beyond the virial analysis of `b2 (core)', we have derived maps of the non-thermal Mach number, $\mathcal{M}_{NT} = \sigma_{NT}/c_{s}$, across the region observed in both \co\ and \cco. The non-thermal velocity dispersion at each pixel is obtained from the observed linewidth via $\sigma_{NT}^{2} = \left(\Delta v / 2\sqrt{2\ln 2}\right)^{2} - k_{B}T_{ex}/m_{obs}$, where $\Delta v$ is the fitted FWHM of the line and $m_{obs}$ is the mass of the observed molecule (29.00 and 30.00 amu for \co\ and \cco, respectively). Linewidths are obtained from Gaussian fits to the spectrum at each pixel over $-$9.5 to $-$2.8 km/s.  Where two velocity components are present along the line of sight (20\% of \co\ pixels, reflecting the two velocity groups identified in Section \ref{sec:data}, and 3\% of \cco\ pixels), the spectrum is fitted with two Gaussians and the brighter component is used, so that the linewidth reflects motions within a single component rather than the separation between them. The isothermal sound speed of the bulk gas is given by $c_{s} = \sqrt{k_{B}T_{ex}/\mu m_{H}}$, adopting a mean molecular weight per free particle of $\mu$ = 2.33 \citep{Kauffmann2008}. As elsewhere in this work (Section \ref{sec:FCRAO}), $T_{ex}$ is taken pixel-by-pixel from the EXFC \tco\ data; since \tco\ is optically thick, its excitation temperature provides the closest available proxy to the kinetic temperature of the gas, which is the quantity required by $c_{s}$. Over the mapped region $T_{ex}$ ranges from 6.1 to 18.2 K, with a median of 12.2 K, corresponding to a median sound speed of 0.208 km/s. Within the velocity range of G148.04+0.11, the ratio of the EXFC \tco\ and \co\ brightness temperatures imply a median \tco\ optical depth of $\sim$30, assuming a $^{12}$C/$^{13}$C ratio of 77 \citep{Milam2005}.The highest values of $T_{ex}$ are found in the brightest \co\ clumps rather than at the cloud edges, consistent with variations in beam filling and excitation rather than local heating.

The measured mass within this region, derived using the pixel-by-pixel \co/\cco\ abundance ratio determined in Section~\ref{sec:Virial}, is 4.0 $\pm$ 1.1~\msol, giving a virial ratio of 2.6 $\pm$ 0.7. We conclude that the observed core is consistent with being marginally bound, with a virial ratio consistent within the uncertainties with $\alpha_{vir} \approx$ 2, the approximate threshold between gravitationally bound and unbound states \citep{Kauffmann2013}. Gravitational collapse therefore cannot be ruled out, although the absence of an associated embedded source suggests that any collapse has not yet progressed to star formation. Additional support from magnetic fields or external pressure may also be present.

We have also estimated virial ratios for the seven \co\ structures, following the same approach. The effective radius of each structure is $R_{\rm eff} = \sqrt{A/\pi}$, where $A$ is the area over which its mass is measured (as established by $Astrodendro$). The observed velocity dispersion is the intensity-weighted quadrature sum of the mean dispersion from Gaussian fits to the individual \co\ spectra and the pixel-to-pixel variation of the fitted centroid velocities. The thermal correction uses the mean $T_{ex}$ over each structure (11.7--14.2 K). Results are presented in Table \ref{tbl:Masses} and error estimates combine mass uncertainties with bootstrap estimates of the uncertainty in $\sigma_{obs}$. The three structures associated with \cco\ emission (`B', `E' and `F') have virial ratios of 1.7--2.0, at or just below the threshold for gravitational boundedness, $\alpha_{vir} \approx$ 2, and are therefore formally bound, whereas `A', `C', `D' and `G' have virial ratios of 4--13 and are not bound by self-gravity alone. Such structures may nevertheless be confined by the pressure of the surrounding medium \citep{Bertoldi1992}, or may be transient. These conclusions are not sensitive to the treatment of the velocity components - where two components are present along the line of sight, only the brighter is used. We note that the virial mass assumed here, $M_v = 5R\sigma_{tot}^2/G$, applies to a uniform sphere, for the more elongated structures, particularly `C' and `D', the virial ratios are indicative only.
We further measure velocity gradients across each \co\ structure by fitting a plane to the intensity-weighted map of fitted centroid velocities. These gradients are 0.2--0.8 km/s/pc (with that of `G' unconstrained), corresponding to velocity differences of $\leq$0.6 km/s across each structure, comparable to or smaller than their velocity dispersions. These values are typical of the gradients found in dense cores and clumps \citep{Goodman1993}.

\begin{figure}
    \centering
    \includegraphics[width=0.49\textwidth]{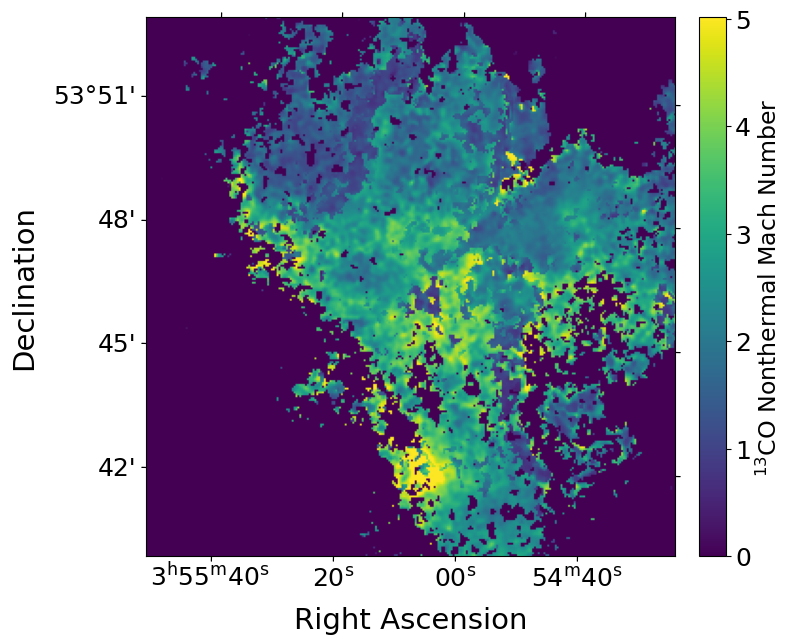}
    \includegraphics[width=0.49\textwidth]{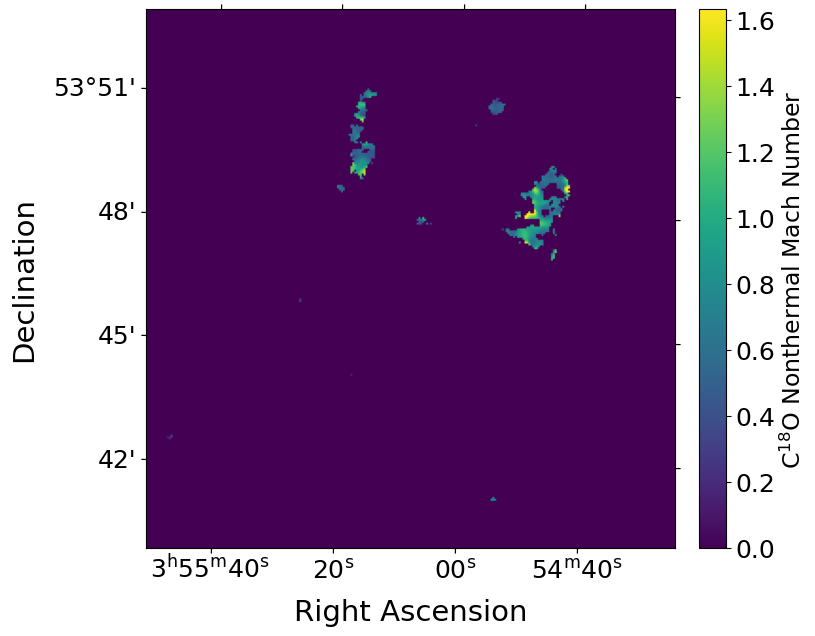}
    \caption{Non-thermal Mach number maps of G148.04+0.11 derived from \co\ (left, 9\arcsec\ resolution) and \cco\ (right, 12\arcsec\ resolution). In both cases the excitation temperature has been taken pixel-by-pixel from the EXFC \tco\ data (Section \ref{sec:FCRAO}).}
    \label{fig:MachNo}
\end{figure}

We find median non-thermal Mach numbers of 2.4 in \co\ (interquartile range 1.8 - 3.1) and 0.7 in \cco\ (interquartile range 0.6 - 0.9), with median non-thermal velocity dispersions of 0.52 and 0.16 km/s, respectively (Fig. \ref{fig:MachNo}). The \co-traced gas is therefore supersonic, with 96\% of pixels exhibiting $\mathcal{M}_{NT} >$ 1, whereas the denser gas traced by \cco\ is predominantly subsonic, with only 19\% of pixels exhibiting $\mathcal{M}_{NT} >$ 1. This transition from supersonic motions in the more extended gas to subsonic motions in the dense condensations is consistent with \cco\ probing denser, more quiescent material in which turbulent motions have dissipated, as found for dense cores more generally \citep{Goodman1998}. The \co\ values fall below the median \co-derived Mach number of $\sim$3.6 reported for the broader PGCC population by \citet{Yuebin2026}, discussed further in Section \ref{sec:disc}. We note that their analysis likewise anchors $T_{ex}$ to a \tco-derived excitation temperature, so this difference does not arise from a methodological inconsistency between the two studies. The moderate degree of supersonic turbulence found here is consistent with the picture developed in Section \ref{sec:disc} of a cold, quiescent cloud that has not yet reached the stage of localised gravitational collapse.

\section{Discussion}
\label{sec:disc}
Our observations show a combination of filamentary and clumpy structures across the region we have designated as G148.04+0.11, with denser, clump-like condensations embedded within more extended, lower-density \co\ emission. Such hierarchical, filamentary structure is common to molecular clouds generally \citep{Andre2014} and does not by itself establish the evolutionary state of a cloud. Our data are consistent with a scenario in which filaments channel material onto clumps that may subsequently fragment \citep{Andre2010,Andre2014}, but do not require it.

The virial analysis (Section \ref{sec:Virial}) shows that the three structures associated with \cco\ emission (`B', `E' and `F') and the dense `b2' core, are close to the threshold for gravitational boundedness and could therefore collapse in the absence of additional support, e.g. from magnetic fields. The remaining structures (`A', `C', `D' and `G') are not gravitationally bound and, unless confined by external pressure \citep{Bertoldi1992}, are likely transient or dispersing. The velocity gradients across the structures are modest and typical of dense clumps and cores. We therefore find no velocity structure indicative of collapse, although our data are not suited to detecting infall directly, which requires line asymmetries in an optically thick tracer at the resolution of the clumps. The largest-scale velocity structure seen in our observations is the $\sim$1.5 km/s offset between the blue-shifted structures (`B' and `F') and the red-shifted structures (`C' and `D'; Section \ref{sec:data}), which may arise from converging flows or from overlapping sub-clouds -  our data cannot distinguish between these possibilities.

Multiple observations indicate that G148.04+0.11 is not in a state of core-collapse/star formation: the absence of embedded infrared sources clearly associated with the dense gas (Section \ref{sec:Embedded}), cold dust temperatures (Section \ref{sec:dust}) and the predominantly subsonic motions of the \cco-traced gas (Section \ref{sec:Virial}). We therefore regard G148.04+0.11 as a cold, quiescent cloud without current star formation, in which the densest clumps are close to gravitational boundedness. Whether it will go on to form stars, or whether its lower-density structures will disperse cannot be determined from the present data.

G148.04+0.11 sits within a broader population whose typical properties have recently been characterised by the TOP-SCOPE CO survey of 2008 PGCCs \citep{Yuebin2026}, which finds these clumps to be cold (median dust temperature $\sim$13 K) but dynamically active, with widespread supersonic non-thermal motions (median \co-derived Mach number $\sim$3.6) that increase mildly with dust-derived column density. Our own data place G148.04+0.11 firmly at the cold end of this distribution: The total integrated column densities of the \co\ structures `A' - `G' arise almost entirely from PPMAP slices below 18.4 K (Section \ref{sec:dust}).

The non-thermal Mach numbers derived in Section \ref{sec:Virial}, with medians of 2.4 in \co\ and 0.7 in \cco, are systematically lower than the \citet{Yuebin2026} population median. Since \citet{Yuebin2026} likewise anchor their excitation temperatures to optically thick \tco\ emission, we believe our comparison is consistent in methodology and therefore interpret G148.04+0.11 as supersonically turbulent but quiescent with respect to the overall PGCC population, with the dense gas traced by \cco\ being predominantly subsonic. This is consistent with the marginally bound state we find for the dense `b2' core (virial ratio of 2.6 $\pm$ 0.7, Section \ref{sec:Virial}), and with a picture in which PGCC substructure at this stage is close to the threshold for gravitational collapse, with turbulence or external pressure providing comparable support.

\begin{figure}
    \centering
    \includegraphics[width=1.0\textwidth]{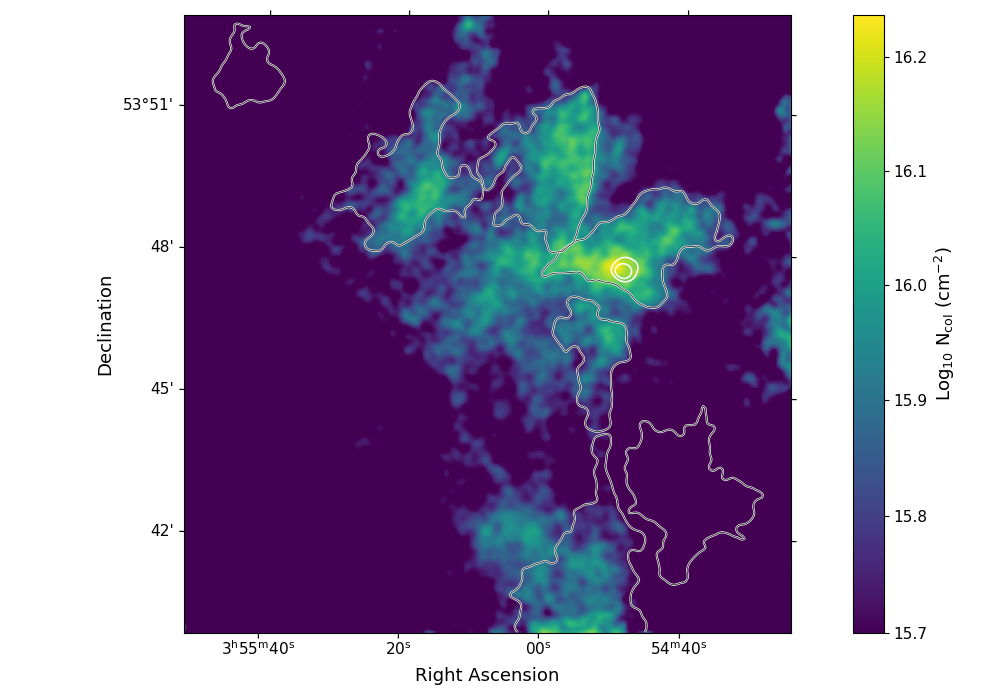}
    \caption{Image of the log of \co\ column density, overlaid with contours of \cco\ column density at levels of log$_{10}$(N$_{\mathrm{col}}$) = 15.03, 15.10, 15.16, 15.23 and 15.30}
    \label{fig:Ncol}
\end{figure}

\citet{Yuebin2026} report that $\bigchi_{CO}$ increases with dust-derived column density, a trend interpreted as evidence of enhanced CO freeze-out in colder, denser gas. Our finding that detectable \cco\ emission is confined to only three of the seven identified \co\ structures (`B', `E' and `F'; Section \ref{sec:data}), despite comparable \co\ column densities being present across the cloud (Fig. \ref{fig:Ncol}), is qualitatively consistent with this depletion picture. However, a quantitative comparison is beyond what can be determined from the present data. Dust continuum coverage of G148.04+0.11 are available (Section \ref{sec:dust}). However, as established in Section \ref{sec:FCRAO}, there are distinct velocity components present in the gas observed here. As such, visible continuum emission cannot be definitively associated with either of the two kinematically distinct components along the line-of-sight. Dust-derived column densities cannot be cleanly apportioned between the sources we have identified here and other, likely background, material. A PPMAP temperature decomposition offers a partial mitigation, in that the coldest slices preferentially trace the gas associated with our \co\ structures (Section \ref{sec:dust}), but it does not isolate G148.04+0.11 unambiguously enough to support a quantitative comparison of $\bigchi_{CO}$ using dust-derived column density.

\section{Summary}
\label{sec:summary}
We present new \co\ and \cco\ (J=1$\rightarrow$0) mapping observations of the Planck Galactic Cold Clump G148.04+0.11, obtained with the Argus array on the Green Bank Telescope at 9\arcsec\ and 12\arcsec\ resolution, respectively. The principal results are summarised as follows.

\begin{enumerate}

\item G148.04+0.11 shows a broad, low-level of \co\ emission. From this, seven \co\ structures have been identified via $Astrodendro$, which we have labelled `A' through `G'. `A' and `G' are relatively weak and isolated, while `B', `C', `D', `E' and `F' form two stronger branches of the dendrogram. `C' and `D' appear associated with filamentary structure towards the southern edge of the region, while `B', `E' and `F' are less elongated and show associated \cco\ emission, within which we identify four denser \cco\ substructures (`b1', `b2', `e1' and `f1').

\item The \co\ moment 1 map reveals a broad kinematic division of the cloud into a blue-shifted group (`B' and `F', peaking near $-$6.75 km/s) and a red-shifted, more filamentary group (`C' and `D', peaking near $-$5.0 km/s), with `E' at an intermediate velocity and `A'/`G' comprising more diffuse, intermediate-velocity material. This is consistent with the upper half of the mapped region showing clumpy structure while the lower half is predominantly filamentary.

\item Comparison with the wider-field, lower-resolution \tco\ and \co\ data from the Exeter-FCRAO Galactic Plane Survey (Section \ref{sec:FCRAO}) reveals that our mapped region represents two, kinematically distinct velocity components in \co, of which our observations show emission over a range of roughly $-$7.0 to $-$4.5 km/s. At velocities around $-$33.6 km/s, emission from the Perseus Arm ($\sim$2.05 kpc) is seen in the EXFC data. This line-of-sight confusion prevents a robust, quantitative association between observed dust emission and the G148.04+0.11 cloud itself.

\item Adopting a distance of 1.327 kpc \citep{Guo2021}, consistent with our own independent extinction-based estimate, we derive a total \co-traced mass of 651~\msol\ for the mapped region, of which the $Astrodendro$ leaf structures account for 280.0~\msol\ (43\%). The \co/\cco\ abundance ratio, derived for the `b1', `b2', `e1' and `f1' regions, ranges from 6.0 -- 16.0 (mean 9.4), broadly consistent with the Galactocentric prediction of \citet{Wilson1994}.

\item A virial analysis of the compact and symmetric \cco\ `b2 (core)' gives a virial mass of 10.3~\msol\ against a measured mass of 4.0~\msol\ (virial ratio 2.6 $\pm$ 0.7), indicating that this core is marginally bound and that gravitational collapse cannot be ruled out, though it may additionally be supported by magnetic fields. Of the seven \co\ structures, those associated with \cco\ emission (`B', `E' and `F') are at or just below the threshold for gravitational boundedness, $\alpha_{vir} \approx$ 2 (virial ratios of 1.7--2.0), and are therefore formally bound, while `A', `C', `D' and `G' are unbound and/or pressure confined (virial ratios of 4--13). Velocity gradients across the structures are modest (0.2--0.8 km/s/pc), with no velocity structure indicative of collapse.

\item Pixel-by-pixel non-thermal Mach number maps, constructed using excitation temperatures derived from the optically thick EXFC \tco\ data and linewidths from Gaussian fits, show the \co-traced gas to be supersonically turbulent, with a median $\mathcal{M}_{NT}$ of 2.4, while the dense gas traced by \cco\ is predominantly subsonic (median $\mathcal{M}_{NT}$ of 0.7). The \co\ value sits below the median \co-derived Mach number ($\sim$3.6) reported for the broader PGCC population by \citet{Yuebin2026}. We interpret G148.04+0.11 as dynamically quiescent relative to the wider PGCC population, consistent with the marginally bound state found for the `b2 (core)'.

\end{enumerate}

Taken together, these results indicate a cold, quiescent cloud without current star formation, in which the clumps associated with \cco\ emission and the dense `b2 (core)' are close to gravitational boundedness, while the lower-density structures are unbound and/or pressure-confined. The data are consistent with an early stage of filament-fed mass assembly and fragmentation, but do not exclude the dispersal of parts of the cloud.

\section{Acknowledgements}
The authors thank an anonymous referee for a detailed and thorough review, which has led to significant improvements in the paper.

The authors wish to thank the operators and staff of the Green Bank Telescope for their assistance in the acquisition of the data presented in this work.

Data and observations from multiple observatories were utilised in preparing this work - \\
$Herschel$ is an ESA space observatory with science instruments provided by European-led Principal Investigator consortia and with important participation from NASA.\\
The James Clerk Maxwell Telescope is operated by the East Asian Observatory on behalf of Academia Sinica Institute of Astronomy and Astrophysics and the National Astronomical Research Institute of Thailand. Additional funding support is provided by the Science and Technology Facilities Council of the United Kingdom and participating universities and organizations in the United Kingdom and Canada. Additional funds for the construction of SCUBA-2 were provided by the Canada Foundation for Innovation.\\

This research made use of $Astrodendro$, a Python package to compute dendrograms of Astronomical data (http://www.dendrograms.org/).

The authors acknowledge the use of generative AI tools in the preparation of this manuscript. According to the GAIDeT taxonomy, the following tasks were delegated to GAI tools under full human supervision:
\begin{itemize}
    \item Drafting small sections of manuscript text, subsequently revised by the authors
    \item Identifying and correcting \latex/BibTeX errors, including cross-references and citation formatting
    \item Reviewing citation completeness and suggesting supporting references, subsequently verified by the authors against the original literature
    \item Assistance with debugging of Python data-reduction and analysis scripts
\end{itemize}
Claude (Anthropic, Claude Sonnet 4.6) was used for manuscript text, \latex/BibTeX correction, and citation review. ChatGPT (OpenAI; specific model version not recorded, as the default freely-available model was used over several years of this project) was used to write Python data-handling code and to locate literature references for CO analysis. All AI-generated content was reviewed, verified, and edited by the authors, who take full responsibility for the accuracy and scientific integrity of the final manuscript. AI tools are not listed as authors and bear no responsibility for the work presented.
\\

\bibliography{G148Paper}{}

\begin{thebibliography}{}
\expandafter\ifx\csname natexlab\endcsname\relax\def\natexlab#1{#1}\fi
\providecommand{\url}[1]{\href{#1}{#1}}
\providecommand{\dodoi}[1]{doi:~\href{http://doi.org/#1}{\nolinkurl{#1}}}
\providecommand{\doeprint}[1]{\href{http://ascl.net/#1}{\nolinkurl{http://ascl.net/#1}}}
\providecommand{\doarXiv}[1]{\href{https://arxiv.org/abs/#1}{\nolinkurl{https://arxiv.org/abs/#1}}}

\bibitem[{{Andr{\'e}} {et~al.}(2014){Andr{\'e}}, {Di Francesco},
  {Ward-Thompson}, {Inutsuka}, {Pudritz}, \& {Pineda}}]{Andre2014}
{Andr{\'e}}, P., {Di Francesco}, J., {Ward-Thompson}, D., {et~al.} 2014, in
  Protostars and Planets VI, ed. H.~{Beuther}, R.~S. {Klessen}, C.~P.
  {Dullemond}, \& T.~{Henning}, 27--51

\bibitem[{{Andr{\'e}} {et~al.}(2010){Andr{\'e}}, {Men'shchikov}, {Bontemps},
  {K{\"o}nyves}, {Motte}, {Schneider}, {Didelon}, {Minier}, {Saraceno},
  {Ward-Thompson}, {di Francesco}, {White}, {Molinari}, {Testi}, {Abergel},
  {Griffin}, {Henning}, {Royer}, {Mer{\'\i}n}, {Vavrek}, {Attard},
  {Arzoumanian}, {Wilson}, {Ade}, {Aussel}, {Baluteau}, {Benedettini},
  {Bernard}, {Blommaert}, {Cambr{\'e}sy}, {Cox}, {di Giorgio}, {Hargrave},
  {Hennemann}, {Huang}, {Kirk}, {Krause}, {Launhardt}, {Leeks}, {Le Pennec},
  {Li}, {Martin}, {Maury}, {Olofsson}, {Omont}, {Peretto}, {Pezzuto}, {Prusti},
  {Roussel}, {Russeil}, {Sauvage}, {Sibthorpe}, {Sicilia-Aguilar}, {Spinoglio},
  {Waelkens}, {Woodcraft}, \& {Zavagno}}]{Andre2010}
{Andr{\'e}}, P., {Men'shchikov}, A., {Bontemps}, S., {et~al.} 2010, \aap, 518,
  L102, \dodoi{10.1051/0004-6361/201014666}

\bibitem[{{Arzoumanian} {et~al.}(2013){Arzoumanian}, {Andr{\'e}}, {Peretto}, \&
  {K{\"o}nyves}}]{Arzoumanian2013}
{Arzoumanian}, D., {Andr{\'e}}, P., {Peretto}, N., \& {K{\"o}nyves}, V. 2013,
  \aap, 553, A119, \dodoi{10.1051/0004-6361/201220822}

\bibitem[{{Bergin} \& {Tafalla}(2007)}]{Bergin2007}
{Bergin}, E.~A., \& {Tafalla}, M. 2007, \araa, 45, 339,
  \dodoi{10.1146/annurev.astro.45.071206.100404}

\bibitem[{{Bertoldi} \& {McKee}(1992)}]{Bertoldi1992}
{Bertoldi}, F., \& {McKee}, C.~F. 1992, \apj, 395, 140, \dodoi{10.1086/171638}

\bibitem[{{B{\'e}thermin} {et~al.}(2010){B{\'e}thermin}, {Dole}, {Beelen}, \&
  {Aussel}}]{Bethermin2010}
{B{\'e}thermin}, M., {Dole}, H., {Beelen}, A., \& {Aussel}, H. 2010, \aap, 512,
  A78, \dodoi{10.1051/0004-6361/200913279}

\bibitem[{{Bolatto} {et~al.}(2013){Bolatto}, {Wolfire}, \&
  {Leroy}}]{Bolatto2013}
{Bolatto}, A.~D., {Wolfire}, M., \& {Leroy}, A.~K. 2013, \araa, 51, 207,
  \dodoi{10.1146/annurev-astro-082812-140944}

\bibitem[{{Chen} {et~al.}(2019){Chen}, {Huang}, {Yuan}, {Wang}, {Fan}, {Xiang},
  {Zhang}, {Tian}, \& {Liu}}]{Chen2019}
{Chen}, B.~Q., {Huang}, Y., {Yuan}, H.~B., {et~al.} 2019, \mnras, 483, 4277,
  \dodoi{10.1093/mnras/sty3341}

\bibitem[{{Cutri} {et~al.}(2003){Cutri}, {Skrutskie}, {van Dyk}, {Beichman},
  {Carpenter}, {Chester}, {Cambresy}, {Evans}, {Fowler}, {Gizis}, {Howard},
  {Huchra}, {Jarrett}, {Kopan}, {Kirkpatrick}, {Light}, {Marsh}, {McCallon},
  {Schneider}, {Stiening}, {Sykes}, {Weinberg}, {Wheaton}, {Wheelock}, \&
  {Zacarias}}]{Cutri2003}
{Cutri}, R.~M., {Skrutskie}, M.~F., {van Dyk}, S., {et~al.} 2003, {2MASS All
  Sky Catalog of point sources.} (IRSA)

\bibitem[{{Dobashi}(2011)}]{Dobashi2011}
{Dobashi}, K. 2011, \pasj, 63, S1, \dodoi{10.1093/pasj/63.sp1.S1}

\bibitem[{{Dobashi} {et~al.}(2005){Dobashi}, {Uehara}, {Kandori}, {Sakurai},
  {Kaiden}, {Umemoto}, \& {Sato}}]{Dobashi2005}
{Dobashi}, K., {Uehara}, H., {Kandori}, R., {et~al.} 2005, \pasj, 57, S1,
  \dodoi{10.1093/pasj/57.sp1.S1}

\bibitem[{{Eden} {et~al.}(2019){Eden}, {Liu}, {Kim}, {Juvela}, {Liu},
  {Tatematsu}, {Francesco}, {Wang}, {Wu}, {Thompson}, {Fuller}, {Li},
  {Ristorcelli}, {Kang}, {Hirano}, {Johnstone}, {Lin}, {He}, {Koch},
  {Sanhueza}, {Qin}, {Zhang}, {Goldsmith}, {Evans}, {Yuan}, {Zhang}, {White},
  {Choi}, {Lee}, {Toth}, {Mairs}, {Yi}, {Tang}, {Soam}, {Peretto}, {Samal},
  {Fich}, {Parsons}, {Malinen}, {Bendo}, {Rivera-Ingraham}, {Liu},
  {Wouterloot}, {Li}, {Qian}, {Rawlings}, {Rawlings}, {Feng}, {Wang}, {Li},
  {Liu}, {Luo}, {Marston}, {Pattle}, {Pelkonen}, {Rigby}, {Zahorecz}, {Zhang},
  {B{\H{o}}gner}, {Aikawa}, {Akhter}, {Alina}, {Bell}, {Bernard}, {Blain},
  {Bronfman}, {Byun}, {Chapman}, {Chen}, {Chen}, {Chen}, {Chen}, {Chen},
  {Chrysostomou}, {Chu}, {Chung}, {Cornu}, {Cosentino}, {Cunningham}, {Demyk},
  {Drabek-Maunder}, {Doi}, {Eswaraiah}, {Falgarone}, {Feh{\'e}r}, {Fraser},
  {Friberg}, {Garay}, {Ge}, {Gear}, {Greaves}, {Guan}, {Harvey-Smith},
  {Hasegawa}, {He}, {Henkel}, {Hirota}, {Holland}, {Hughes}, {Jarken}, {Ji},
  {Jimenez-Serra}, {Kang}, {Kawabata}, {Kim}, {Kim}, {Kim}, {Kim}, {Koo},
  {Kwon}, {Kuan}, {Lacaille}, {Lai}, {Lee}, {Lee}, {Lee}, {Li}, {Lo}, {Lopez},
  {Lu}, {Lyo}, {Mardones}, {McGehee}, {Meng}, {Montier}, {Montillaud}, {Moore},
  {Morata}, {Moriarty-Schieven}, {Ohashi}, {Pak}, {Park}, {Paladini}, {Pech},
  {Qiu}, {Ren}, {Richer}, {Sakai}, {Shang}, {Shinnaga}, {Stamatellos}, {Tang},
  {Traficante}, {Vastel}, {Viti}, {Walsh}, {Wang}, {Wang}, {Ward-Thompson},
  {Whitworth}, {Wilson}, {Xu}, {Yang}, {Yuan}, {Yuan}, {Zavagno}, {Zhang},
  {Zhang}, {Zhang}, {Zhou}, {Zhou}, {Zhu}, \& {Zuo}}]{Eden2019}
{Eden}, D.~J., {Liu}, T., {Kim}, K.-T., {et~al.} 2019, \mnras, 485, 2895,
  \dodoi{10.1093/mnras/stz574}

\bibitem[{{Edenhofer} {et~al.}(2024){Edenhofer}, {Zucker}, {Frank}, {Saydjari},
  {Speagle}, {Finkbeiner}, \& {En{\ss}lin}}]{Edenhofer2024}
{Edenhofer}, G., {Zucker}, C., {Frank}, P., {et~al.} 2024, \aap, 685, A82,
  \dodoi{10.1051/0004-6361/202347628}

\bibitem[{{Elia} {et~al.}(2021){Elia}, {Merello}, {Molinari}, {Schisano},
  {Zavagno}, {Russeil}, {M{\`e}ge}, {Martin}, {Olmi}, {Pestalozzi}, {Plume},
  {Ragan}, {Benedettini}, {Eden}, {Moore}, {Noriega-Crespo}, {Paladini},
  {Palmeirim}, {Pezzuto}, {Pilbratt}, {Rygl}, {Schilke}, {Strafella}, {Tan},
  {Traficante}, {Baldeschi}, {Bally}, {Giorgio}, {Fiorellino}, {Liu}, {Piazzo},
  \& {Polychroni}}]{Elia2021}
{Elia}, D., {Merello}, M., {Molinari}, S., {et~al.} 2021, \mnras, 504, 2742,
  \dodoi{10.1093/mnras/stab1038}

\bibitem[{{Goldsmith} \& {Langer}(1999)}]{Goldsmith1999}
{Goldsmith}, P.~F., \& {Langer}, W.~D. 1999, \apj, 517, 209,
  \dodoi{10.1086/307195}

\bibitem[{{Goodman} {et~al.}(1998){Goodman}, {Barranco}, {Wilner}, \&
  {Heyer}}]{Goodman1998}
{Goodman}, A.~A., {Barranco}, J.~A., {Wilner}, D.~J., \& {Heyer}, M.~H. 1998,
  \apj, 504, 223, \dodoi{10.1086/306045}

\bibitem[{{Goodman} {et~al.}(1993){Goodman}, {Benson}, {Fuller}, \&
  {Myers}}]{Goodman1993}
{Goodman}, A.~A., {Benson}, P.~J., {Fuller}, G.~A., \& {Myers}, P.~C. 1993,
  \apj, 406, 528, \dodoi{10.1086/172465}

\bibitem[{Green(2018)}]{Green2018}
Green, G.~M. 2018, Journal of Open Source Software, 3, 695,
  \dodoi{10.21105/joss.00695}

\bibitem[{{Green} {et~al.}(2019){Green}, {Schlafly}, {Zucker}, {Speagle}, \&
  {Finkbeiner}}]{Green2019}
{Green}, G.~M., {Schlafly}, E., {Zucker}, C., {Speagle}, J.~S., \&
  {Finkbeiner}, D. 2019, \apj, 887, 93, \dodoi{10.3847/1538-4357/ab5362}

\bibitem[{{Guo} {et~al.}(2021){Guo}, {Chen}, {Li}, {Huang}, {Yang}, {Li},
  {Sun}, \& {Liu}}]{Guo2021}
{Guo}, H.~L., {Chen}, B.~Q., {Li}, G.~X., {et~al.} 2021, \mnras, 500, 3743,
  \dodoi{10.1093/mnras/staa3478}

\bibitem[{{Holland} {et~al.}(1999){Holland}, {Robson}, {Gear}, {Cunningham},
  {Lightfoot}, {Jenness}, {Ivison}, {Stevens}, {Ade}, {Griffin}, {Duncan},
  {Murphy}, \& {Naylor}}]{Holland1999}
{Holland}, W.~S., {Robson}, E.~I., {Gear}, W.~K., {et~al.} 1999, \mnras, 303,
  659, \dodoi{10.1046/j.1365-8711.1999.02111.x}

\bibitem[{{Johnstone} \& {Bally}(2006)}]{Johnstone2006}
{Johnstone}, D., \& {Bally}, J. 2006, \apj, 653, 383, \dodoi{10.1086/508852}

\bibitem[{{Kauffmann} {et~al.}(2008){Kauffmann}, {Bertoldi}, {Bourke}, {Evans},
  \& {Lee}}]{Kauffmann2008}
{Kauffmann}, J., {Bertoldi}, F., {Bourke}, T.~L., {Evans}, II, N.~J., \& {Lee},
  C.~W. 2008, \aap, 487, 993, \dodoi{10.1051/0004-6361:200809481}

\bibitem[{{Kauffmann} {et~al.}(2013){Kauffmann}, {Pillai}, \&
  {Goldsmith}}]{Kauffmann2013}
{Kauffmann}, J., {Pillai}, T., \& {Goldsmith}, P.~F. 2013, \apj, 779, 185,
  \dodoi{10.1088/0004-637X/779/2/185}

\bibitem[{{Kim} {et~al.}(2017){Kim}, {Lee}, {Liu}, {Kim}, {Wu}, {Tatematsu},
  {Liu}, {JCMT Large Program {\textquotedblleft}SCOPE{\textquotedblright}
  Collaboration}, \& {TRAO Key Science Program
  {\textquotedblleft}TOP{\textquotedblright} Collaboration}}]{Kim2017}
{Kim}, J., {Lee}, J.-E., {Liu}, T., {et~al.} 2017, \apjs, 231, 9,
  \dodoi{10.3847/1538-4365/aa7b31}

\bibitem[{{Liu} {et~al.}(2013){Liu}, {Wu}, \& {Zhang}}]{Liu2013}
{Liu}, T., {Wu}, Y., \& {Zhang}, H. 2013, \apjl, 775, L2,
  \dodoi{10.1088/2041-8205/775/1/L2}

\bibitem[{{Lynds}(1965)}]{Lynds1965}
{Lynds}, B.~T. 1965, \apjs, 12, 163, \dodoi{10.1086/190123}

\bibitem[{{MacLaren} {et~al.}(1988){MacLaren}, {Richardson}, \&
  {Wolfendale}}]{MacLaren1988}
{MacLaren}, I., {Richardson}, K.~M., \& {Wolfendale}, A.~W. 1988, \apj, 333,
  821, \dodoi{10.1086/166791}

\bibitem[{{Mangum} \& {Shirley}(2015)}]{Mangum2015}
{Mangum}, J.~G., \& {Shirley}, Y.~L. 2015, \pasp, 127, 266,
  \dodoi{10.1086/680323}

\bibitem[{{Marsh} {et~al.}(2017){Marsh}, {Whitworth}, {Lomax}, {Ragan},
  {Becciani}, {Cambr{\'e}sy}, {Di Giorgio}, {Eden}, {Elia}, {Kacsuk},
  {Molinari}, {Palmeirim}, {Pezzuto}, {Schneider}, {Sciacca}, \&
  {Vitello}}]{Marsh2017}
{Marsh}, K.~A., {Whitworth}, A.~P., {Lomax}, O., {et~al.} 2017, \mnras, 471,
  2730, \dodoi{10.1093/mnras/stx1723}

\bibitem[{{Marton} {et~al.}(2017){Marton}, {Calzoletti}, {Perez Garcia},
  {Kiss}, {Paladini}, {Altieri}, {Sanchez Portal}, {Kidger}, \& {the Herschel
  Point Source Catalogue Working Group}}]{Marton2017}
{Marton}, G., {Calzoletti}, L., {Perez Garcia}, A.~M., {et~al.} 2017, arXiv
  e-prints, arXiv:1705.05693, \dodoi{10.48550/arXiv.1705.05693}

\bibitem[{{McDowell}(1988)}]{McDowell1988}
{McDowell}, R.~S. 1988, \jcp, 88, 356, \dodoi{10.1063/1.454608}

\bibitem[{{Milam} {et~al.}(2005){Milam}, {Savage}, {Brewster}, {Ziurys}, \&
  {Wyckoff}}]{Milam2005}
{Milam}, S.~N., {Savage}, C., {Brewster}, M.~A., {Ziurys}, L.~M., \& {Wyckoff},
  S. 2005, \apj, 634, 1126, \dodoi{10.1086/497123}

\bibitem[{{Molinari} {et~al.}(2011){Molinari}, {Schisano}, {Faustini},
  {Pestalozzi}, {di Giorgio}, \& {Liu}}]{Molinari2011}
{Molinari}, S., {Schisano}, E., {Faustini}, F., {et~al.} 2011, \aap, 530, A133,
  \dodoi{10.1051/0004-6361/201014752}

\bibitem[{{Molinari} {et~al.}(2010){Molinari}, {Swinyard}, {Bally}, {Barlow},
  {Bernard}, {Martin}, {Moore}, {Noriega-Crespo}, {Plume}, {Testi}, {Zavagno},
  {Abergel}, {Ali}, {Andr{\'e}}, {Baluteau}, {Benedettini}, {Bern{\'e}},
  {Billot}, {Blommaert}, {Bontemps}, {Boulanger}, {Brand}, {Brunt}, {Burton},
  {Campeggio}, {Carey}, {Caselli}, {Cesaroni}, {Cernicharo}, {Chakrabarti},
  {Chrysostomou}, {Codella}, {Cohen}, {Compiegne}, {Davis}, {de Bernardis}, {de
  Gasperis}, {Di Francesco}, {di Giorgio}, {Elia}, {Faustini}, {Fischera},
  {Fukui}, {Fuller}, {Ganga}, {Garcia-Lario}, {Giard}, {Giardino}, {Glenn},
  {Goldsmith}, {Griffin}, {Hoare}, {Huang}, {Jiang}, {Joblin}, {Joncas},
  {Juvela}, {Kirk}, {Lagache}, {Li}, {Lim}, {Lord}, {Lucas}, {Maiolo},
  {Marengo}, {Marshall}, {Masi}, {Massi}, {Matsuura}, {Meny}, {Minier},
  {Miville-Desch{\^e}nes}, {Montier}, {Motte}, {M{\"u}ller}, {Natoli}, {Neves},
  {Olmi}, {Paladini}, {Paradis}, {Pestalozzi}, {Pezzuto}, {Piacentini},
  {Pomar{\`e}s}, {Popescu}, {Reach}, {Richer}, {Ristorcelli}, {Roy}, {Royer},
  {Russeil}, {Saraceno}, {Sauvage}, {Schilke}, {Schneider-Bontemps},
  {Schuller}, {Schultz}, {Shepherd}, {Sibthorpe}, {Smith}, {Smith},
  {Spinoglio}, {Stamatellos}, {Strafella}, {Stringfellow}, {Sturm}, {Taylor},
  {Thompson}, {Tuffs}, {Umana}, {Valenziano}, {Vavrek}, {Viti}, {Waelkens},
  {Ward-Thompson}, {White}, {Wyrowski}, {Yorke}, \& {Zhang}}]{Molinari2010}
{Molinari}, S., {Swinyard}, B., {Bally}, J., {et~al.} 2010, \pasp, 122, 314,
  \dodoi{10.1086/651314}

\bibitem[{{Molinari} {et~al.}(2016){Molinari}, {Schisano}, {Elia},
  {Pestalozzi}, {Traficante}, {Pezzuto}, {Swinyard}, {Noriega-Crespo}, {Bally},
  {Moore}, {Plume}, {Zavagno}, {di Giorgio}, {Liu}, {Pilbratt}, {Mottram},
  {Russeil}, {Piazzo}, {Veneziani}, {Benedettini}, {Calzoletti}, {Faustini},
  {Natoli}, {Piacentini}, {Merello}, {Palmese}, {Del Grande}, {Polychroni},
  {Rygl}, {Polenta}, {Barlow}, {Bernard}, {Martin}, {Testi}, {Ali},
  {Andr{\'e}}, {Beltr{\'a}n}, {Billot}, {Carey}, {Cesaroni}, {Compi{\`e}gne},
  {Eden}, {Fukui}, {Garcia-Lario}, {Hoare}, {Huang}, {Joncas}, {Lim}, {Lord},
  {Martinavarro-Armengol}, {Motte}, {Paladini}, {Paradis}, {Peretto},
  {Robitaille}, {Schilke}, {Schneider}, {Schulz}, {Sibthorpe}, {Strafella},
  {Thompson}, {Umana}, {Ward-Thompson}, \& {Wyrowski}}]{Molinari2016}
{Molinari}, S., {Schisano}, E., {Elia}, D., {et~al.} 2016, \aap, 591, A149,
  \dodoi{10.1051/0004-6361/201526380}

\bibitem[{{Motte} {et~al.}(1998){Motte}, {Andre}, \& {Neri}}]{Motte1998}
{Motte}, F., {Andre}, P., \& {Neri}, R. 1998, \aap, 336, 150

\bibitem[{{Mottram} \& {Brunt}(2010)}]{Mottram2010}
{Mottram}, J.~C., \& {Brunt}, C.~M. 2010, in Astronomical Society of the
  Pacific Conference Series, Vol. 438, The Dynamic Interstellar Medium: A
  Celebration of the Canadian Galactic Plane Survey, ed. R.~{Kothes}, T.~L.
  {Landecker}, \& A.~G. {Willis}, 98

\bibitem[{{Myers} \& {Benson}(1983)}]{Myers1983}
{Myers}, P.~C., \& {Benson}, P.~J. 1983, \apj, 266, 309, \dodoi{10.1086/160780}

\bibitem[{{Pineda} {et~al.}(2008){Pineda}, {Caselli}, \&
  {Goodman}}]{Pineda2008}
{Pineda}, J.~E., {Caselli}, P., \& {Goodman}, A.~A. 2008, \apj, 679, 481,
  \dodoi{10.1086/586883}

\bibitem[{{Planck Collaboration} {et~al.}(2011){Planck Collaboration}, {Ade},
  {Aghanim}, {Arnaud}, {Ashdown}, {Aumont}, {Baccigalupi}, {Balbi}, {Banday},
  {Barreiro}, {Bartlett}, {Battaner}, {Benabed}, {Beno{\^\i}t}, {Bernard},
  {Bersanelli}, {Bhatia}, {Bonaldi}, {Bonavera}, {Bond}, {Borrill}, {Bouchet},
  {Bucher}, {Burigana}, {Butler}, {Cabella}, {Cantalupo}, {Cappellini},
  {Cardoso}, {Carvalho}, {Catalano}, {Cay{\'o}n}, {Challinor}, {Chamballu},
  {Chary}, {Chen}, {Chiang}, {Chiang}, {Christensen}, {Clements}, {Colombi},
  {Couchot}, {Coulais}, {Crill}, {Cuttaia}, {Danese}, {Davis}, {de Bernardis},
  {de Rosa}, {de Zotti}, {Delabrouille}, {Delouis}, {D{\'e}sert}, {Dickinson},
  {Diego}, {Dolag}, {Dole}, {Donzelli}, {Dor{\'e}}, {D{\"o}rl}, {Douspis},
  {Dupac}, {Efstathiou}, {En{\ss}lin}, {Eriksen}, {Finelli}, {Forni},
  {Fosalba}, {Frailis}, {Franceschi}, {Galeotta}, {Ganga}, {Giard},
  {Giraud-H{\'e}raud}, {Gonz{\'a}lez-Nuevo}, {G{\'o}rski}, {Gratton},
  {Gregorio}, {Gruppuso}, {Haissinski}, {Hansen}, {Harrison}, {Helou},
  {Henrot-Versill{\'e}}, {Hern{\'a}ndez-Monteagudo}, {Herranz}, {Hildebrandt},
  {Hivon}, {Hobson}, {Holmes}, {Hornstrup}, {Hovest}, {Hoyland},
  {Huffenberger}, {Huynh}, {Jaffe}, {Jones}, {Juvela}, {Keih{\"a}nen},
  {Keskitalo}, {Kisner}, {Kneissl}, {Knox}, {Kurki-Suonio}, {Lagache},
  {L{\"a}hteenm{\"a}ki}, {Lamarre}, {Lasenby}, {Laureijs}, {Lawrence}, {Leach},
  {Leahy}, {Leonardi}, {Le{\'o}n-Tavares}, {Leroy}, {Lilje},
  {Linden-V{\o}rnle}, {L{\'o}pez-Caniego}, {Lubin}, {Mac{\'\i}as-P{\'e}rez},
  {MacTavish}, {Maffei}, {Maggio}, {Maino}, {Mandolesi}, {Mann}, {Maris},
  {Marleau}, {Marshall}, {Mart{\'\i}nez-Gonz{\'a}lez}, {Masi}, {Massardi},
  {Matarrese}, {Matthai}, {Mazzotta}, {McGehee}, {Meinhold}, {Melchiorri},
  {Melin}, {Mendes}, {Mennella}, {Mitra}, {Miville-Desch{\^e}nes}, {Moneti},
  {Montier}, {Morgante}, {Mortlock}, {Munshi}, {Murphy}, {Naselsky}, {Natoli},
  {Netterfield}, {N{\o}rgaard-Nielsen}, {Noviello}, {Novikov}, {Novikov},
  {O'Dwyer}, {Osborne}, {Pajot}, {Paladini}, {Partridge}, {Pasian},
  {Patanchon}, {Pearson}, {Perdereau}, {Perotto}, {Perrotta}, {Piacentini},
  {Piat}, {Piffaretti}, {Plaszczynski}, {Platania}, {Pointecouteau}, {Polenta},
  {Ponthieu}, {Poutanen}, {Pratt}, {Pr{\'e}zeau}, {Prunet}, {Puget}, {Rachen},
  {Reach}, {Rebolo}, {Reinecke}, {Renault}, {Ricciardi}, {Riller},
  {Ristorcelli}, {Rocha}, {Rosset}, {Rowan-Robinson}, {Rubi{\~n}o-Mart{\'\i}n},
  {Rusholme}, {Sajina}, {Sandri}, {Santos}, {Savini}, {Schaefer}, {Scott},
  {Seiffert}, {Shellard}, {Smoot}, {Starck}, {Stivoli}, {Stolyarov},
  {Sudiwala}, {Sunyaev}, {Sygnet}, {Tauber}, {Tavagnacco}, {Terenzi},
  {Toffolatti}, {Tomasi}, {Torre}, {Tristram}, {Tuovinen}, {T{\"u}rler},
  {Umana}, {Valenziano}, {Valiviita}, {Varis}, {Vielva}, {Villa}, {Vittorio},
  {Wade}, {Wandelt}, {White}, {Wilkinson}, {Yvon}, {Zacchei}, \&
  {Zonca}}]{Planck2011}
{Planck Collaboration}, {Ade}, P.~A.~R., {Aghanim}, N., {et~al.} 2011, \aap,
  536, A7, \dodoi{10.1051/0004-6361/201116474}

\bibitem[{{Planck Collaboration} {et~al.}(2016){Planck Collaboration}, {Ade},
  {Aghanim}, {Arnaud}, {Ashdown}, {Aumont}, {Baccigalupi}, {Banday},
  {Barreiro}, {Bartolo}, {Battaner}, {Benabed}, {Beno{\^\i}t},
  {Benoit-L{\'e}vy}, {Bernard}, {Bersanelli}, {Bielewicz}, {Bonaldi},
  {Bonavera}, {Bond}, {Borrill}, {Bouchet}, {Boulanger}, {Bucher}, {Burigana},
  {Butler}, {Calabrese}, {Catalano}, {Chamballu}, {Chiang}, {Christensen},
  {Clements}, {Colombi}, {Colombo}, {Combet}, {Couchot}, {Coulais}, {Crill},
  {Curto}, {Cuttaia}, {Danese}, {Davies}, {Davis}, {de Bernardis}, {de Rosa},
  {de Zotti}, {Delabrouille}, {D{\'e}sert}, {Dickinson}, {Diego}, {Dole},
  {Donzelli}, {Dor{\'e}}, {Douspis}, {Ducout}, {Dupac}, {Efstathiou}, {Elsner},
  {En{\ss}lin}, {Eriksen}, {Falgarone}, {Fergusson}, {Finelli}, {Forni},
  {Frailis}, {Fraisse}, {Franceschi}, {Frejsel}, {Galeotta}, {Galli}, {Ganga},
  {Giard}, {Giraud-H{\'e}raud}, {Gjerl{\o}w}, {Gonz{\'a}lez-Nuevo},
  {G{\'o}rski}, {Gratton}, {Gregorio}, {Gruppuso}, {Gudmundsson}, {Hansen},
  {Hanson}, {Harrison}, {Helou}, {Henrot-Versill{\'e}},
  {Hern{\'a}ndez-Monteagudo}, {Herranz}, {Hildebrandt}, {Hivon}, {Hobson},
  {Holmes}, {Hornstrup}, {Hovest}, {Huffenberger}, {Hurier}, {Jaffe}, {Jaffe},
  {Jones}, {Juvela}, {Keih{\"a}nen}, {Keskitalo}, {Kisner}, {Knoche}, {Kunz},
  {Kurki-Suonio}, {Lagache}, {Lamarre}, {Lasenby}, {Lattanzi}, {Lawrence},
  {Leonardi}, {Lesgourgues}, {Levrier}, {Liguori}, {Lilje}, {Linden-V{\o}rnle},
  {L{\'o}pez-Caniego}, {Lubin}, {Mac{\'\i}as-P{\'e}rez}, {Maggio}, {Maino},
  {Mandolesi}, {Mangilli}, {Marshall}, {Martin}, {Mart{\'\i}nez-Gonz{\'a}lez},
  {Masi}, {Matarrese}, {Mazzotta}, {McGehee}, {Melchiorri}, {Mendes},
  {Mennella}, {Migliaccio}, {Mitra}, {Miville-Desch{\^e}nes}, {Moneti},
  {Montier}, {Morgante}, {Mortlock}, {Moss}, {Munshi}, {Murphy}, {Naselsky},
  {Nati}, {Natoli}, {Netterfield}, {N{\o}rgaard-Nielsen}, {Noviello},
  {Novikov}, {Novikov}, {Oxborrow}, {Paci}, {Pagano}, {Pajot}, {Paladini},
  {Paoletti}, {Pasian}, {Patanchon}, {Pearson}, {Pelkonen}, {Perdereau},
  {Perotto}, {Perrotta}, {Pettorino}, {Piacentini}, {Piat}, {Pierpaoli},
  {Pietrobon}, {Plaszczynski}, {Pointecouteau}, {Polenta}, {Pratt},
  {Pr{\'e}zeau}, {Prunet}, {Puget}, {Rachen}, {Reach}, {Rebolo}, {Reinecke},
  {Remazeilles}, {Renault}, {Renzi}, {Ristorcelli}, {Rocha}, {Rosset},
  {Rossetti}, {Roudier}, {Rubi{\~n}o-Mart{\'\i}n}, {Rusholme}, {Sandri},
  {Santos}, {Savelainen}, {Savini}, {Scott}, {Seiffert}, {Shellard}, {Spencer},
  {Stolyarov}, {Sudiwala}, {Sunyaev}, {Sutton}, {Suur-Uski}, {Sygnet},
  {Tauber}, {Terenzi}, {Toffolatti}, {Tomasi}, {Tristram}, {Tucci}, {Tuovinen},
  {Umana}, {Valenziano}, {Valiviita}, {Van Tent}, {Vielva}, {Villa}, {Wade},
  {Wandelt}, {Wehus}, {Yvon}, {Zacchei}, \& {Zonca}}]{PGCC2016}
---. 2016, \aap, 594, A28, \dodoi{10.1051/0004-6361/201525819}

\bibitem[{{Reid} {et~al.}(2016){Reid}, {Dame}, {Menten}, \&
  {Brunthaler}}]{Reid2016}
{Reid}, M.~J., {Dame}, T.~M., {Menten}, K.~M., \& {Brunthaler}, A. 2016, \apj,
  823, 77, \dodoi{10.3847/0004-637X/823/2/77}

\bibitem[{{Reid} {et~al.}(2009){Reid}, {Menten}, {Zheng}, {Brunthaler},
  {Moscadelli}, {Xu}, {Zhang}, {Sato}, {Honma}, {Hirota}, {Hachisuka}, {Choi},
  {Moellenbrock}, \& {Bartkiewicz}}]{Reid2009}
{Reid}, M.~J., {Menten}, K.~M., {Zheng}, X.~W., {et~al.} 2009, \apj, 700, 137,
  \dodoi{10.1088/0004-637X/700/1/137}

\bibitem[{{Reid} {et~al.}(2014){Reid}, {Menten}, {Brunthaler}, {Zheng}, {Dame},
  {Xu}, {Wu}, {Zhang}, {Sanna}, {Sato}, {Hachisuka}, {Choi}, {Immer},
  {Moscadelli}, {Rygl}, \& {Bartkiewicz}}]{Reid2014}
{Reid}, M.~J., {Menten}, K.~M., {Brunthaler}, A., {et~al.} 2014, \apj, 783,
  130, \dodoi{10.1088/0004-637X/783/2/130}

\bibitem[{{Reid} {et~al.}(2019){Reid}, {Menten}, {Brunthaler}, {Zheng}, {Dame},
  {Xu}, {Li}, {Sakai}, {Wu}, {Immer}, {Zhang}, {Sanna}, {Moscadelli}, {Rygl},
  {Bartkiewicz}, {Hu}, {Quiroga-Nu{\~n}ez}, \& {van Langevelde}}]{Reid2019}
---. 2019, \apj, 885, 131, \dodoi{10.3847/1538-4357/ab4a11}

\bibitem[{{Ripple} {et~al.}(2013){Ripple}, {Heyer}, {Gutermuth}, {Snell}, \&
  {Brunt}}]{Ripple2013}
{Ripple}, F., {Heyer}, M.~H., {Gutermuth}, R., {Snell}, R.~L., \& {Brunt},
  C.~M. 2013, \mnras, 431, 1296, \dodoi{10.1093/mnras/stt247}

\bibitem[{{Rivera-Ingraham} {et~al.}(2016){Rivera-Ingraham}, {Ristorcelli},
  {Juvela}, {Montillaud}, {Men'shchikov}, {Malinen}, {Pelkonen}, {Marston},
  {Martin}, {Pagani}, {Paladini}, {Paradis}, {Ysard}, {Ward-Thompson},
  {Bernard}, {Marshall}, {Montier}, \& {T{\'o}th}}]{Rivera-Ingraham2016}
{Rivera-Ingraham}, A., {Ristorcelli}, I., {Juvela}, M., {et~al.} 2016, \aap,
  591, A90, \dodoi{10.1051/0004-6361/201526263}

\bibitem[{Robitaille {et~al.}(2019)Robitaille, Rice, Beaumont, Ginsburg,
  MacDonald, \& Rosolowsky}]{Robitaille2019}
Robitaille, T., Rice, T., Beaumont, C., {et~al.} 2019, astrodendro:
  Astronomical data dendrogram creator, Astrophysics Source Code Library,
  ascl:1907.016

\bibitem[{{Rosolowsky} {et~al.}(2008){Rosolowsky}, {Pineda}, {Kauffmann}, \&
  {Goodman}}]{Rosolowsky2008}
{Rosolowsky}, E.~W., {Pineda}, J.~E., {Kauffmann}, J., \& {Goodman}, A.~A.
  2008, \apj, 679, 1338, \dodoi{10.1086/587685}

\bibitem[{{Roueff} {et~al.}(2021){Roueff}, {Gerin}, {Gratier}, {Levrier},
  {Pety}, {Gaudel}, {Goicoechea}, {Orkisz}, {de Souza Magalhaes}, {Vono},
  {Bardeau}, {Bron}, {Chanussot}, {Chainais}, {Guzman}, {Hughes},
  {Kainulainen}, {Languignon}, {Le Bourlot}, {Le Petit}, {Liszt}, {Marchal},
  {Miville-Desch{\^e}nes}, {Peretto}, {Roueff}, \& {Sievers}}]{Roueff2021}
{Roueff}, A., {Gerin}, M., {Gratier}, P., {et~al.} 2021, \aap, 645, A26,
  \dodoi{10.1051/0004-6361/202037776}

\bibitem[{{Sawada} {et~al.}(2008){Sawada}, {Ikeda}, {Sunada}, {Kuno},
  {Kamazaki}, {Morita}, {Kurono}, {Koura}, {Abe}, {Kawase}, {Maekawa},
  {Horigome}, \& {Yanagisawa}}]{Sawada2008}
{Sawada}, T., {Ikeda}, N., {Sunada}, K., {et~al.} 2008, \pasj, 60, 445,
  \dodoi{10.1093/pasj/60.3.445}

\bibitem[{{Seale} {et~al.}(2012){Seale}, {Looney}, {Wong}, {Ott}, {Klein}, \&
  {Pineda}}]{Seale2012}
{Seale}, J.~P., {Looney}, L.~W., {Wong}, T., {et~al.} 2012, \apj, 751, 42,
  \dodoi{10.1088/0004-637X/751/1/42}

\bibitem[{{Sharpless}(1959)}]{Sharpless1959}
{Sharpless}, S. 1959, \apjs, 4, 257, \dodoi{10.1086/190049}

\bibitem[{{Sieth} {et~al.}(2014){Sieth}, {Devaraj}, {Voll}, {Church},
  {Gawande}, {Cleary}, {Readhead}, {Kangaslahti}, {Samoska}, {Gaier},
  {Goldsmith}, {Harris}, {Gundersen}, {Frayer}, {White}, {Egan}, \&
  {Reeves}}]{Sieth2014}
{Sieth}, M., {Devaraj}, K., {Voll}, P., {et~al.} 2014, in Society of
  Photo-Optical Instrumentation Engineers (SPIE) Conference Series, Vol. 9153,
  Millimeter, Submillimeter, and Far-Infrared Detectors and Instrumentation for
  Astronomy VII, ed. W.~S. {Holland} \& J.~{Zmuidzinas}, 91530P

\bibitem[{{Tafalla} {et~al.}(2004){Tafalla}, {Myers}, {Caselli}, \&
  {Walmsley}}]{Tafalla2004}
{Tafalla}, M., {Myers}, P.~C., {Caselli}, P., \& {Walmsley}, C.~M. 2004, \aap,
  416, 191, \dodoi{10.1051/0004-6361:20031704}

\bibitem[{{Wilson} \& {Rood}(1994)}]{Wilson1994}
{Wilson}, T.~L., \& {Rood}, R. 1994, \araa, 32, 191,
  \dodoi{10.1146/annurev.aa.32.090194.001203}

\bibitem[{{Wu} {et~al.}(2012){Wu}, {Liu}, {Meng}, {Li}, {Qin}, \&
  {Ju}}]{Wu2012}
{Wu}, Y., {Liu}, T., {Meng}, F., {et~al.} 2012, \apj, 756, 76,
  \dodoi{10.1088/0004-637X/756/1/76}

\bibitem[{{Yuan} {et~al.}(2016){Yuan}, {Wu}, {Liu}, {Zhang}, {Zeng Li}, {Liu},
  {Meng}, {Chen}, {Hu}, \& {Wang}}]{Yuan2016}
{Yuan}, J., {Wu}, Y., {Liu}, T., {et~al.} 2016, \apj, 820, 37,
  \dodoi{10.3847/0004-637X/820/1/37}

\bibitem[{{Yuebin} {et~al.}(2026){Yuebin}, {Esimbek}, {Liu}, {Baan}, {Liu},
  {Kim}, {Wu}, {Tang}, {Zhou}, {Li}, {He}, {Kang}, {Ma}, \&
  {Zhou}}]{Yuebin2026}
{Yuebin}, Y., {Esimbek}, J., {Liu}, T., {et~al.} 2026, \aj, 172, 146,
  \dodoi{10.3847/1538-3881/ae8cff}

\bibitem[{{Zhang} {et~al.}(2018){Zhang}, {Liu}, {Yuan}, {Sanhueza},
  {Traficante}, {Li}, {Li}, {Tatematsu}, {Wang}, {Lee}, {Samal}, {Eden},
  {Marston}, {Liu}, {Zhou}, {Li}, {Koch}, {Xu}, {Wu}, {Juvela}, {Zhang},
  {Alina}, {Goldsmith}, {T{\'o}th}, {Wang}, \& {Kim}}]{Zhang2018}
{Zhang}, C.-P., {Liu}, T., {Yuan}, J., {et~al.} 2018, \apjs, 236, 49,
  \dodoi{10.3847/1538-4365/aac513}

\bibitem[{{Zhang} {et~al.}(2016){Zhang}, {Wu}, {Liu}, \& {Meng}}]{Zhang2016}
{Zhang}, T., {Wu}, Y., {Liu}, T., \& {Meng}, F. 2016, \apjs, 224, 43,
  \dodoi{10.3847/0067-0049/224/2/43}

\end{thebibliography}
\bibliographystyle{aasjournal}

\end{document}